\documentclass[12pt]{article}
\usepackage{amsfonts,amsmath,amssymb,epsf}
\usepackage{graphicx,color}
 \def\ep{{\epsilon}}

 \def\frac#1#2{{#1\over #2}}

 \def\s{\sqrt}

 \def\al{\alpha'}
 \def\de{\partial}

 \def\f {\frac}
 \def\ti{\tilde}
 \def\ap{\alpha}

 \def\ddd{\cdot\cdot\cdot}
 \def\no{\nonumber \\}
 \def\la{\langle}
 \def\lb{\rangle}
 \def\ep{\epsilon}

 \def\vp{\varphi}

 \def\ep{{\epsilon}}

 \def\frac#1#2{{#1\over #2}}

 \def\s{\sqrt}

\def\be{\begin{equation}}
\def\ee{\end{equation}}
\def\ba{\begin{eqnarray}}
\def\ea{\end{eqnarray}}

\begin{document}
\begin{titlepage}
\thispagestyle{empty}
\begin{flushright}
NSF-KITP-06-31 \\
KUNS-2021\\
hep-th/0605073 \\
May, 2006
\end{flushright}

\bigskip

\begin{center}
\noindent{\Large \textbf{Aspects of Holographic Entanglement Entropy}}\\
\vspace{2cm} \noindent{Shinsei Ryu\footnote{e-mail:
sryu@kitp.ucsb.edu} and Tadashi Takayanagi\footnote{e-mail:
         takayana@gauge.scphys.kyoto-u.ac.jp}$^,$~\footnote{Moved from
           Kavli Institute for Theoretical Physics,
         University of California, Santa Barbara, CA 93106, USA}}\\
\vspace{1cm}

${}^1$ {\it Kavli Institute for Theoretical Physics,
         University of California, \\ Santa Barbara, CA 93106, USA}

\vspace{3mm}
 ${}^2$
 {\it  Department of Physics, Kyoto University, Kyoto 606-8502, Japan }

\end{center}

\begin{abstract}
This is an extended version of our short report hep-th/0603001,
where a holographic interpretation of entanglement entropy in
conformal field theories is proposed from AdS/CFT correspondence. In
addition to a concise review of relevant recent progresses of
entanglement entropy and details omitted in the earlier letter, this
paper includes the following several new results : We give a more
direct derivation of our claim which relates the entanglement
entropy with the minimal area surfaces in the AdS$_3/$CFT$_2$ case
as well as some further discussions on higher dimensional cases.
Also the relation between the entanglement entropy and central
charges in 4D conformal field theories is examined. We check that
the logarithmic part of the 4D entanglement entropy computed in the
CFT side agrees with the AdS$_5$ result at least under a specific
condition. Finally we estimate the entanglement entropy of massive
theories in generic dimensions by making use of our proposal.

\end{abstract}
\end{titlepage}

\newpage

\begin{scriptsize}
\tableofcontents
\end{scriptsize}

\newpage

\section{Introduction}
\label{intro} \setcounter{equation}{0} \hspace{5mm} When we study
properties of a given quantum field theory (QFT), it is common to
first investigate behaviors of correlation functions of local
operators in the theory. However, properties of non-local quantities
are equally important, especially for understanding of its quantum
mechanical phase structure. One basic such example of non-local
physical quantities is the Wilson loop operators in gauge theories,
which is a very useful order parameter of confinement.

In a more generic class of QFTs, we can instead consider a quantity
called entanglement entropy (or geometric entropy). This is defined
as the von Neumann entropy $S_A$ when we `trace out' (or smear out)
degrees of freedom inside a $d$-dimensional space-like submanifold
$B$ in a given $d+1$ dimensional QFT. Its complement is denoted by
the submanifold $A$. It measures how a given quantum system is
entangled or strongly correlated. Intuitively we can also say that
this is the entropy for an observer in $A$ who is not accessible to
$B$ as the information is lost by the smearing out in region $B$.

As its name suggests, we expect that the entanglement entropy is
directly related to the degrees of freedom. Indeed, the entanglement
entropy is proportional to the central charge in two dimensional
conformal field theories (2D CFTs) as first pointed out in
\cite{HLW}.  Recently, this property was also confirmed in
\cite{Cardy} in which a general prescription of computing the
entropy in 2D CFTs is given.
 Also in the mass perturbed CFTs (massive QFTs) the same
conclusion holds \cite{Vidal03, Latorre04, Cardy}. Furthermore, as
we will discuss later, the similar statement is also true in 4D
CFTs. The entropy is related to the 4D central charges.

In higher dimensional (more than two dimensional) QFTs, it is not
easy to compute the entanglement entropy for arbitrary submanifolds
$A$ even in free field theories. Motivated by this, we would like to
consider a holographic estimation of the quantity by applying
AdS/CFT correspondence (or duality) \cite{Maldacena,adsreview}. We
can find pioneering works \cite{HMS,MBH} that discuss related issues
from slightly different viewpoints. Recently, the authors of the
present paper proposed a holographic computation of entanglement
entropy in CFTs from the AdS/CFT \cite{RuTa}. This reduces the
complicated quantum calculation in QFTs to a classical differential
geometrical computation.

The AdS/CFT correspondence relates a $d+2$ dimensional AdS space
(AdS$_{d+2}$) to a $d+1$ dimensional CFT (CFT$_{d+1}$), which is
sitting at the boundary of the AdS$_{d+2}$. The claim is that the
entropy $S_A$ in a $d+1$ dimensional CFT can be determined from the
$d$ dimensional minimal surface $\gamma_A$ whose boundary is given
by the $d-1$ dimensional manifold $\de \gamma_A=\de A$.  The entropy
is given by applying the Bekenstein-Hawking entropy formula to the
area of the minimal surface $\gamma_A$ as if $\gamma_A$ were an
event horizon. This is motivated by the idea of the entropy bound
\cite{holography,BiSu,Bousso} and by the similarity between the
black hole horizons and the minimal surface $\gamma_A$. They become
equivalent in the special cases such as those in AdS black holes
\cite{RuTa} and in black holes of brane-world \cite{Emparan}, as the
minimal surfaces wrap the horizons (see also \cite{HMS,MBH,Einhorn}
for earlier related discussions\footnote{Also refer to \cite{entan}
for recent arguments on the relation between the entanglement
entropy in three qubit systems and the entropy of BPS black holes,
based on similarities of their symmetries, though this does not
seems to be related to our issues directly.}). In \cite{RuTa} we
have shown that our proposal, when applied to the lowest dimensional
AdS$_3/$CFT$_2$ example, correctly reproduces the known results of
entanglement entropy in 2D CFT. Also it is easy to see that the
Bekenstein-Hawking like formula is consistent with the known `area
law' in entanglement entropy \cite{Bombelli,Srednicki} for the CFTs
(also QFTs) in any dimensions.

In the present paper we would like to study the entanglement entropy
in higher dimensional CFTs, especially CFT$_4$ from both the CFT and
gravity sides. In particular, we find the computations of the
logarithmic term from both sides agree at least when the second
fundamental form of $\de A$ embedded in the $d$ dimensional space
vanishes. In addition, we present a review of the required knowledge
of the entanglement entropy in conformal field theories and the
details of our short report.

We would also like to mention recent interests in entanglement
entropy in condensed matter physics. One of main foci in modern
condensed matter physics is to understand quantum phases of matter
which are beyond the Ginzburg-Landau paradigm. Many-body
wavefunctions of quantum ground states in these phases look
featureless when one looks at correlation functions of local
operators; They cannot be characterized by classical order
parameters of some kind. Indeed, they should be distinguished by
their pattern of entanglement rather than their pattern of symmetry
breaking \cite{Wen89}. Thus, the entanglement entropy is potentially
useful to characterize these exotic phases.

Indeed, this idea has been pushed extensively in recent couple of
years for several 1D quantum systems. It has been revealed that
several quantum phases in 1D spin chains, including Haldane phases,
can be distinguished by different scaling of the entanglement
entropy. See, for example,
\cite{Vidal03, Latorre04, Peschel04, Jin03, Its05} and
references in \cite{Cardy}.

For higher dimensional condensed matter systems,
there has been not many works in this direction yet.
Recently,
the entanglement entropy was applied
for so-called topological phases
in 2+1 D
\cite{Kitaev05, Levin05}.
Typically, these phases
have a finite gap and
are accompanied
by many exotic features such as
fractionalization of quantum numbers,
non-Abelian statistics of quasi-particles,
topological degeneracy, etc.
They can be also useful
fault tolerant quantum computations.

On the other hand, unconventional quantum liquid phases with gapless
excitations, such as gapless spin liquid phases, seem to be, at
least at present, more difficult to characterize in higher
dimensions. Our results from AdS/CFT correspondence can be useful to
study these gapless spin liquid states (some of these phases have
been suspected to be described by a relativistic gauge field theory
of some sort \cite{Wen89}) .

The organization of the present paper is as follows. In section
\ref{basics} we present a review of definition and basic properties
of entanglement entropy. Section \ref{2D CFT} is devoted to
computations of entanglement entropy in 2D CFTs. In section
\ref{(d+1)D CFT} we first summarize the known facts on entanglement
entropy in higher dimensional CFTs and perform explicit computations
especially for 4D CFTs. Next we relate the central charges in a
given 4D CFT to its entanglement entropy. In section
\ref{holographic} we present our proposal of holographic
computations of entanglement entropy from AdS/CFT. We also give an
explicit proof of this claim in AdS$_3/$CFT$_2$ based on the
well-known relation \cite{ADSGKP,ADSWitten} and discuss its
extension to higher dimensional cases. Based on our proposal, in
section \ref{AdS3/CFT2}, we compute the entanglement entropy in 2D
CFTs from the AdS$_3$ side and find agreements. Higher dimensional
cases are considered in section \ref{AdS d+2/CFT d+1} where we
compute the entropy from the analysis of AdS$_{d+2}$ spaces. We
compare it with the CFT results especially for AdS$_5/$CFT$_4$ case
and find an agreement under a specific condition for simplification.
We also estimate entanglement entropy in massive or non-conformal
theories. In section \ref{conclusion} we summarize our results and
discuss future problems.

\section{Basics of Entanglement Entropy}
\label{basics}
\setcounter{equation}{0}
\hspace{5mm}
We start with a review of basic ideas and properties of entanglement entropy.

\subsection{Definition of Entanglement Entropy}
\label{def EE}
\hspace{5mm} Consider  a quantum mechanical system with many degrees
of freedom such as spin chains. More generally, we can consider
arbitrary lattice models or QFTs including CFTs.
We put the system at zero temperature and
then the total quantum system is described by the pure ground state
$|\Psi\lb$. We assume no degeneracy of the ground state.
Then, the density
matrix is that of the pure state \be \rho_{tot}=|\Psi\rangle \langle
\Psi|. \label{pure}\ee The von Neumann entropy of the total system
is clearly zero $ S_{tot}= -\mathrm{tr}\, \rho_{tot} \log
\rho_{tot}=0$.

Next we divide the total system into two subsystems $A$ and $B$. In
the spin chain example, we just artificially cut off the chain at
some point and divide the lattice points into two groups. Notice
that physically we do not do anything to the system and the cutting
procedure is an imaginary process. Accordingly the total Hilbert
space can be written as a direct product of two spaces
${\mathcal{H}}_{tot}={\mathcal{H}}_{A}\otimes {\mathcal{H}}_{B}$
corresponding to those of subsystems $A$ and $B$.  The observer who
is only accessible to the subsystem $A$ will feel as if the total
system is described by the reduced density matrix $\rho_A$
\be
\rho_A= \mathrm{tr}_{B}~\rho_{tot},
\ee where the
trace is taken only over the Hilbert space ${\mathcal{H}}_{B}$.

Now we define the entanglement entropy of the
subsystem $A$ as
the von Neumann entropy of the reduced
density matrix $\rho_A$
\begin{eqnarray}
S_A =
- \mathrm{tr}_{A}\,
\rho_{A} \log \rho_{A}.
\label{eq:def entanglement entropy}
\end{eqnarray}
This quantity provides us with a convenient
way to measure how closely entangled (or how ``quantum'') a given
wave function $|\Psi\rangle$ is. Notice also that in time-dependent
backgrounds the density matrix $\rho_{tot}$ and $\rho_A$ are time
dependent as dictated by the von Neumann equation. Thus we need to
specify the time $t=t_0$ when we measure the entropy. In this paper,
we always study static systems and we can neglect this issue.

It is also possible to define the entanglement entropy $S_A(\beta)$
at finite temperature $T=\beta^{-1}$.
This can be done just by replacing (\ref{pure}) with the thermal one
$\rho_{thermal}=e^{-\beta H}$,
where $H$ is the total Hamiltonian.
When $A$ is the total system, $S_A(\beta)$ is clearly
the same as the thermal entropy.

\subsection{Properties}
\label{prop EE}
\hspace{5mm} There are several useful properties which the entanglement
entropy satisfies generally. We consider the zero temperature case.
We summarize some of them as follows:
\begin{itemize}
\item (i) When $B$ is the complement of $A$ as before,
we obtain \be S_A=S_B. \label{ext}
              \ee
        This manifestly shows that the entanglement
        entropy is not an extensive quantity.
              This equality (\ref{ext}) is violated
              at finite temperature.

\item (ii) When $A$ is divided into two submanifolds
$A_1$ and $A_2$, we find

                 \be S_{A_1}+S_{A_2}\geq S_{A}.
                 \label{extt} \ee  This is called subadditivity.

\item (iii)
For any three subsystems $A$, $B$ and $C$
that do not intersect each other,
the following strong subadditivity inequality holds :
\begin{eqnarray}
S_{A+B+C} + S_{B} &\le& S_{A+B}+S_{B+C}.
\label{exttt}
\end{eqnarray}
Equivalently, we can have a more strong
version of (\ref{extt}) as follows
\begin{eqnarray}
S_{A} + S_{B} &\ge& S_{A\cup B}+S_{A\cap B}, \label{extttst}
\end{eqnarray}
for any subsystems $A$ and $B$. When $A$ and $B$ do not intersect
with each other,
this relation is reduced to the subadditivity
(\ref{extt}.)
\end{itemize}

More details of properties of the
entanglement entropy can be found in e.g. \cite{Nielsen-Chuang00}.

\subsection{Entanglement Entropy in QFTs and Area Law}
\label{EE in QFT, Area}
\hspace{5mm}  Consider a QFT on a $d+1$ dimensional manifold
$\mathbb{R}\times N$ , where $\mathbb{R}$ and $N$
 denote the time direction and the $d$ dimensional
 space-like manifold, respectively.
We define the subsystem by a $d$ dimensional submanifold $A\subset
N$ at fixed time $t=t_0$. We call its complement the submanifold
$B$. The boundary of $A$, which is denoted by $\de A$, divides the
manifold $N$ into two submanifolds $A$ and $B$. Then we can define
the entanglement entropy $S_A$ by the previous formula (\ref{eq:def
entanglement entropy}). Sometimes this kind of entropy is called
geometric entropy as it depends on the geometry of the submanifold
$A$. Since the entanglement entropy is always divergent in a
continuum theory
we introduce an
ultraviolet cut off $a$ (or a lattice spacing).
 Then the coefficient in
front of the divergence turns out to be proportional to the area of
the boundary $\de A$ of the subsystem $A$ as first pointed out in
\cite{Bombelli,Srednicki}, \be S_A=\gamma\cdot \f{\mbox{Area}(\de
A)}{a^{d-1}}+\mbox{subleading terms},
 \label{divarea}\ee
where $\gamma$ is a constant which depends on the system. This
behavior can be intuitively understood since the entanglement
between $A$ and $B$ occurs at the boundary $\de A$ most strongly.
This result (\ref{divarea}) was originally found from numerical
computations \cite{Srednicki,Bombelli} and checked in many later
arguments (see e.g. recent works \cite{Eisert, Das, Casiniarea} ).

The simple area law (\ref{divarea}), however, does not always
describe the scaling of the entanglement entropy in generic
situations. As we will discuss in details in the later sections, the
entanglement entropy of 1D quantum systems at criticality scales
logarithmically with respect to the linear size $l$ of $A$, $S_A
\sim \frac{c}{3}\log l/a$ where $c$ is the central charge of the CFT
that describes the critical point. It has been also recently pointed
out that the area law is corrected by a logarithmic factor as $S_A
\propto  (l/a)^{d-1}\log l/a+\ $(subleading terms) for fermionic
systems in the presence of a finite Fermi surface, where $l$ is the
characteristic length scale of the $d-1$ dimensional manifold $\de
A$ \cite{Wolf05, Gioev05, Barthel06, Li06}. Since we mainly consider
relativistic QFTs (without a finite Fermi surface) in this paper,
the area law (\ref{divarea}) applies to our examples for $d\geq 2$
as we will see.

Before we proceed to further analysis of entanglement entropy, it
might be interesting to notice that this area law (\ref{divarea})
looks very similar to the Bekenstein-Hawking entropy (BH entropy)
of black holes
which is proportional to the area of the event horizon
\begin{equation} S_{BH}=\frac{{\rm
Area~of~horizon}}{4G_N}, \label{BHentropy}
\end{equation}
where $G_N$ is the Newton constant. Intuitively, we can regard $S_A$
as the entropy for an observer who is only accessible to the
subsystem $A$ and cannot receive any signals from $B$. In this
sense, the subsystem $B$ is analogous to the inside of a black hole
horizon for an observer sitting in $A$, i.e., outside of the
horizon. Indeed, this similarity was an original motivation of the
entanglement entropy \cite{Bombelli,Srednicki} (earlier related idea
can also be found in \cite{Thooft}). Even though this analogy is not
completely correct as it is, the one-loop quantum correction to the
BH entropy in the presence of matter fields is known to be equal to
the entanglement entropy \cite{SuUg}. This interesting relation is
an important hint to find the holographic dual of the entanglement
entropy discussed later. Indeed, the connection between this
relation and our proposal has been found recently in \cite{Emparan}
by employing the brane-world setup instead of $AdS$ backgrounds.

\section{Entanglement Entropy in $2\mathrm{D}$ CFT}
\label{2D CFT}
\hspace{5mm}
Here we review and slightly extend
existing computations of
entanglement entropy in $(1+1)$ D CFTs.
 The central charge of a given CFT is denoted by $c$. Such
a computation was initiated in \cite{HLW,FPST} and a general
prescription how to calculate the quantity was given in a recent
work \cite{Cardy} (see also \cite{Casini05a}), which we will explain
in an orbifold theoretic manner. We separately discuss this lowest
dimensional CFT since only in this case we can exactly compute the
entropy for general systems at present.

\subsection{How to Compute Entanglement Entropy}
\label{how to EE}

\hspace{5mm}
In order to find the entanglement entropy, we first
evaluate $\mathrm{tr}_A\,\rho_A^n$, differentiate it with
respect to $n$ and finally take the limit $n\to 1$ (remember that
$\rho_A$ is normalized such that $\mathrm{tr}_A\,\rho_A=1$)
\begin{eqnarray}
S_{A}&=&
\lim_{n\to 1}
\frac{\mathrm{tr}_A\,\rho_A^n-1}{1-n}
\label{Tsallis entropy}
\\
&=&
-\frac{\partial}{\partial n}
\mathrm{tr}_A\,\rho_A^n|_{n=1}
=
-\f{\de}{\de n}\log \mathrm{tr}_A~\rho_A^n|_{n=1}.
\label{deri}
\end{eqnarray}
This is called the replica trick. Therefore, what we have to do is
to evaluate $\mathrm{tr}_A~\rho_A^n$ in our 2D system. The first
line of the above definition (\ref{Tsallis entropy}) without taking
the $n\to 1$ limit defines the so-called Tsallis entropy,
$S_{n,\mathrm{Tsallis}} =\frac{\mathrm{tr}_A\,\rho_A^n-1}{1-n} $.
\footnote{ The Tsallis entropy is related to the alpha entropy
(R\'enyi entropy) $S_{\alpha} =\frac{\log
\mathrm{tr}_A\,\rho_A^\alpha}{1-\alpha} $ through $
S_{\alpha,\mathrm{Tsallis}} = \frac{1}{1-\alpha}
[e^{(1-\alpha)S_{\alpha}}-1] $ \cite{Jin03}. The $\alpha\to 1$  and
$\alpha\to \infty$ limits of the alpha entropy give the von Neumann
entropy and the single-copy entanglement entropy, respectively. }

\begin{figure}
\begin{center}
\includegraphics[height=6cm]{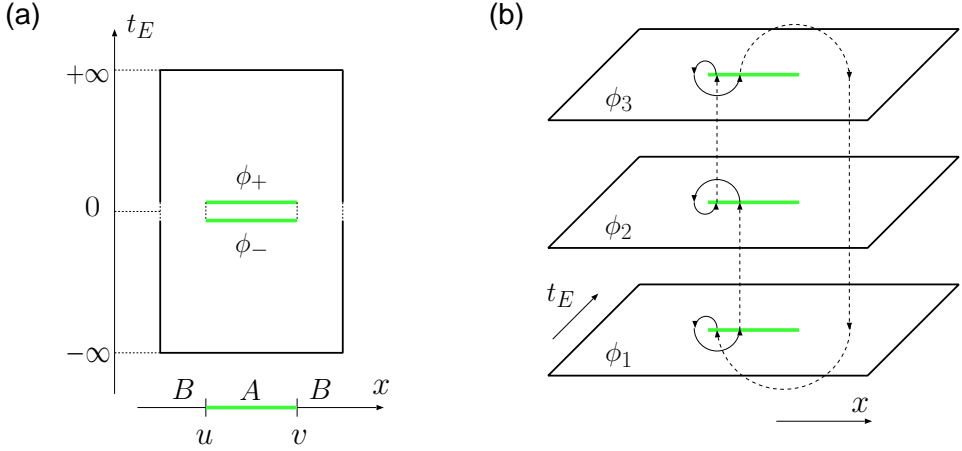}
\end{center}
\caption{
\label{fig: n-riemann.eps}
(a)
The path integral representation of the reduced
density matrix $[\rho_A]_{\phi_+\phi_-}$.
(b)
The $n$-sheeted Riemann surface $\mathcal{R}_n$.
(Here we take $n=3$ for simplicity.)
}
\end{figure}

This can be done in the path-integral formalism as follows. We first
assume that $A$ is the single interval $x\in [u,v]$ at $t_E=0$ in
the flat Euclidean coordinates $(t_E,x)\in \mathbb{R}^{2}$. The
ground state wave function $\Psi$ can be found by path-integrating
from $t_{E}=-\infty$ to $t_{E}=0$ in the Euclidean formalism \be
\Psi\left(\phi_0(x)\right)=
\int^{\phi(t_{E}=0,x)=\phi_0(x)}_{t_{E}=-\infty} D\phi~
e^{-S(\phi)},\ee where $\phi(t_E,x)$ denotes the field which defines
the 2D CFT. The values of the field at the boundary $\phi_0$ depends
on the spacial coordinate $x$. The total density matrix $\rho$ is
given by two copies of the wave function
$[\rho]_{\phi_0\phi_0'}=\Psi(\phi_0)\bar{\Psi}(\phi'_0)$. The
complex conjugate one $\bar{\Psi}$ can be obtained by
path-integrating from $t_{E}=\infty$ to $t_{E}=0$. To obtain the
reduced density matrix $\rho_A$, we need to integrate $\phi_0$ on
$B$ assuming $\phi_{0}(x)=\phi'_0(x)$ when $x\in B$. \be
[\rho_A]_{\phi_+ \phi_-}=(Z_{1})^{-1}\int
^{t_E=\infty}_{t_{E}=-\infty} D\phi~ e^{-S(\phi)}\prod_{x\in
A}\delta\left(\phi(+0,x)-\phi_+(x)\right)\cdot
\delta\left(\phi(-0,x)-\phi_-(x)\right), \label{pathrho}\ee where
$Z_1$ is the vacuum partition function on $\mathbb{R}^{2}$ and we
multiply its inverse in order to normalize $\rho_A$ such that
$\mathrm{tr}_A\,\rho_A=1$. This computation is sketched in Fig.\
\ref{fig: n-riemann.eps} (a).

To find $\mathrm{tr}_A\,\rho_A^n$, we can prepare $n$ copies of
(\ref{pathrho}) \be [\rho_A]_{\phi_{1+}
\phi_{1-}}[\rho_A]_{\phi_{2+} \phi_{2-}}\ddd [\rho_A]_{\phi_{n+}
\phi_{n-}},\ee and take the trace successively. In the path-integral
formalism this is realized by gluing $\{\phi_{i\pm}(x)\}$ as
$\phi_{i-}(x)=\phi_{(i+1)+}(x)$ ($i=1,2,\ddd,n$) and integrating
$\phi_{i+}(x)$. In this way, $\mathrm{tr}_A\,\rho_A^n$ is given in
terms of the path-integral on an $n$-sheeted Riemann surface
${\mathcal R}_n$ (see Fig.\ \ref{fig: n-riemann.eps} (b)) \be
\mathrm{tr}_A\,\rho_A^n=(Z_1)^{-n}\int_{(t_E,x)\in {\mathcal R}_n}
D\phi~ e^{-S(\phi)}\equiv \f{Z_n}{(Z_1)^n}. \label{nsheetp}  \ee

To evaluate the path-integral on ${\mathcal R}_n$, it is useful to
introduce replica fields. Let us first take $n$ disconnected sheets.
The field on each sheet is denoted by
$\phi_{k}(t_E,x)$ ($k=1,2,\ddd,n$). In order to obtain a CFT on the
flat complex plane $\mathbb{C}$ which is equivalent to the present
one on ${\mathcal R}_n$, we impose the twisted boundary conditions
 \be \phi_k(e^{2\pi
i}(w-u))=\phi_{k+1}(w-u),\ \ \ \phi_k(e^{2\pi
i}(w-v))=\phi_{k-1}(w-v), \label{bct} \ee where we employed the
complex coordinate $w=x+it_E$. Equivalently we can regard the
boundary condition (\ref{bct}) as the insertion of two twist
operators $\Phi^{+(k)}_n$ and $\Phi^{-(k)}_n$ at $w=u$ and $w=v$ for
each ($k-$th) sheet. Thus we find \be
\mathrm{tr}_A\,\rho_A^n=\prod_{k=0}^{n-1}
\la\Phi^{+(k)}_n(u)\Phi^{-(k)}_n(v)\lb. \label{densityend} \ee

\subsection{Derivation of Entanglement Entropy in an Infinitely Long System}
\label{deriv EE in inf sys} \hspace{5mm} When $\phi$ is a real
scalar field, this is a non-abelian orbifold. To make the situation
simple, assume that $\phi$ is a complex scalar field. Then we can
diagonalize the boundary condition by defining $n$ new fields
$\ti{\phi}_k=\f{1}{n}\sum_{l=1}^ne^{2\pi il k /n}\phi_l$. They obey
the boundary condition \be \ti{\phi}_k(e^{2\pi i}(w-u))=e^{2\pi
ik/n}\ti{\phi}_{k}(w-u), \ \ \ \ti{\phi}_k(e^{2\pi i}(w-v))=e^{-2\pi
ik/n}\ti{\phi}_{k}(w-v). \label{bctt} \ee Thus in this case we can
conclude that the system is equivalent to $n-$disconnected sheets
with two twist operators $\sigma_{k/n}$ and $\sigma_{-k/n}$ inserted
in the $k-$th sheet for each values of $k$. In the end we find \be
\mathrm{tr}_A\,\rho_A^n=\prod_{k=0}^{n-1}
\la\sigma_{k/n}(u)\sigma_{-k/n}(v)\lb \sim
(u-v)^{-4\sum_{k=0}^{n-1}\Delta_{k/n}}=(u-v)^{-\f{1}{3}(n-1/n)} ,\ee
where $\Delta_{k/n}=-\f{1}{2}\left(\f{k}{n}\right)^2+\f12\f{k}{n}$
is the (chiral) conformal dimension of $\sigma_{k/n}$. When we have
$m$ such complex scalar fields we simply obtain \be
\mathrm{tr}_{A}\,\rho_A^n=\prod_{k=0}^{n-1}
\la\sigma_{k/n}(u)\sigma_{-k/n}(v)\lb \sim (u-v)^{-\f{c}{6}(n-1/n)}
,\label{general}\ee setting the central charge $c=2m$.

To deal with a general CFT with central charge $c$, we need to go
back to the basis (\ref{bct}). The paper \cite{Cardy} showed that
the result (\ref{general}) is generally correct (see also
\cite{Kn}). The argument is roughly as follows. Define the
coordinate $z$ as follows \be
z=\left(\f{w-u}{w-v}\right)^{\f{1}{n}}. \label{conformaltr}
\ee This
maps ${\mathcal{R}}_n$ to the $z$-plane $\mathbb{C}$. In this simple
coordinate system we easily find $\la T(z)\lb_{\mathbb{C}}=0$. Via
Schwartz derivative term in the conformal map we obtain a
non-vanishing value of $\la T(w)\lb_{\mathcal{R}_n}$. From that
result, we can learn that twist operators
$\Phi^{\pm(k)}_n$ in (\ref{densityend})
have conformal dimension $\Delta_n=\f{c}{24}(1-n^{-2})$.
Thus we find the
same result (\ref{general}) for general CFTs as follows from
(\ref{densityend}).

Applying the formula (\ref{deri}) to (\ref{general}), we
find\footnote{Here we neglect a constant term which does not depend
on $l$, $L$ and $a$.} the famous result \cite{HLW} \be S_A
=\f{c}{3}~\log\f{l}{a}, \label{simple} \ee where $a$ is the UV cut
off (or lattice spacing) and we set $l\equiv v-u$.

It is possible to extend the above result to the general case where
$A$ consists of multi intervals \be A=\{w|{\rm Im}\, w=0, {\rm Re}\,
w\in [u_1,v_1]\cup [u_2,v_2]\cup \ddd \cup [u_{N},v_{N}]\}.
\label{multideff}\ee We obtain the value of the trace \cite{Cardy}
\be \mathrm{tr}_A\,\rho_{A_w}^n\sim \left(\f{\prod_{1\leq j<k\leq
N}(u_k-u_j)(v_k-v_j)}{\prod_{j,k=1}^N
(v_k-u_j)}\right)^{\f{c}{6}(n-1/n)}. \label{tracett} \ee Thus the
entanglement entropy is given as follows \cite{Cardy} \be S_A
=\f{c}{3}\sum_{1\leq i,j\leq
N}\log\f{u_i-v_j}{a}-\f{c}{3}\sum_{1\leq i<j\leq
N}\log\f{u_j-u_i}{a}-\f{c}{3}\sum_{1\leq i<j\leq
N}\log\f{v_j-v_i}{a}. \label{multient} \ee

\subsection{Derivation of Entanglement Entropy on a Circle}
\label{deriv EE on circ}
\hspace{5mm} We assume the space direction $x$ is compactified as a
circle of circumference $L$. The system $A$ is defined by the
subsystem $A$ by the union \be A=\{x|x\in [r_1,s_1]\cup
[r_2,s_2]\cup \ddd \cup [r_{N},s_{N}]\}, \label{multib} \ee where we
assume $0\leq r_1<s_1<r_2< s_2< \ddd < r_N<s_N\leq L$.  This
subsystem $A$ is related to the previous one (\ref{multideff}) via
the conformal map \be w=\tan\left(\f{\pi w'}{L}\right).
\label{confrmappp} \ee This maps the previous $n$-sheeted Riemann
surface $w\in {\mathcal R}_n$ to the $n$-sheeted cylinder $w'\in
{\mathcal Cyl}_n$. We find $u_i=\tan\left(\f{\pi r_i}{L}\right)$ and
$v_i=\tan\left(\f{\pi s_i}{L}\right)$.

To compute $\mathrm{tr}_A\,\rho_{A_{w'}}^n$ in
 this cylinder coordinates, we can
apply the conformal transformations (\ref{confrmappp}). This leads
to the extra factor \be
\prod_{i=1}^{N}\left[\f{L}{\pi}\cos\left(\f{\pi
r_i}{L}\right)\cos\left(\f{\pi
s_i}{L}\right)\right]^{-\f{c}{6}(1-n^{-2})}, \label{zerotem} \ee
which should be multiplied with (\ref{tracett}). In this way, the
entanglement entropy is given by
\ba
S_A &=&\f{c}{3}\sum_{1\leq i,j\leq
N}\log\left(\f{L}{\pi a}\sin\left(\f{\pi(r_i-s_j)}{L}\right)\right)
\nonumber \\
&-&
\f{c}{3}\sum_{1\leq i<j\leq N}
\log\left(\f{L}{\pi a}\sin\left(\f{\pi(r_j-r_i)}{L} \right)\right)-
\f{c}{3} \sum_{1\leq i<j\leq N} \log\left(\f{L}{\pi
a}\sin\left(\f{\pi(s_j-s_i)}{L}\right)\right).
\nonumber \\
\label{entamiult} \ea

When we only have one interval with the length $l$,
(\ref{entamiult}) is reduced to the known result \cite{HLW,Cardy}
\be
S_A = \f{c}{3}\cdot\log\left(\f{L}{\pi a}\sin\left(\f{\pi
l}{L}\right)\right). \label{entropyone}
\ee Notice that in the small $l$ limit, (\ref{entropyone})
approaches to (\ref{simple}) as expected.
Also the expression  (\ref{entropyone})
is invariant under the exchange $l\to L-l$
 and thus satisfies the property (\ref{ext}).

\subsection{Derivation of Entanglement Entropy at Finite Temperature}
\label{deriv EE at finit T}
\hspace{5mm} It is also possible to calculate $S_A$ at finite
temperature $T=\beta^{-1}$ when its spacial length is infinite
$L=\infty$. In this case we need to compactify the Euclidean time as
$t_E\sim t_E+\beta$. We can map this system to the previous one
(\ref{multideff}) via the conformal map \be
w=e^{\f{2\pi}{\beta}w'}. \label{confmapfin} \ee  We find
$u_i=e^{\f{2\pi r_i}{\beta}}$ and $v_i=e^{\f{2\pi s_i}{\beta}}$ .
This conformal map leads to the extra factor \be
\prod_{i=1}^{N}\left[\f{\beta}{2\pi}~e^{-\f{\pi}{\beta}(r_i+s_i)}
\right]^{-\f{c}{6}(1-n^{-2})}, \label{fintem} \ee in addition to
(\ref{tracett}).
Thus we obtain $S_A$ as follows
\ba
S_A&=&\f{c}{3}\sum_{1\leq i,j\leq N}\log\left(\f{\beta}{\pi
a}\sinh\left(\f{\pi(r_i-s_j)}{\beta}\right)\right)
\nonumber \\
&-&
\f{c}{3}\sum_{1\leq i<j\leq N}\log\left(\f{\beta}{\pi a}
\sinh\left(\f{\pi(r_j-r_i)}{\beta}\right)\right)
-\f{c}{3}\sum_{1\leq i<j\leq N}\log\left(\f{\beta}{\pi a}
\sinh\left(\f{\pi(s_j-s_i)}{\beta}\right)\right).
\nonumber\\
\label{entamiultem} \ea
 If the subsystem $A$ is a single length $l$ segment, it
becomes the known result \cite{Cardy}
\be S_A=\f{c}{3}\cdot\log\left(\f{\beta}{\pi
a}\sinh\left(\f{\pi l}{\beta}\right)\right).
\label{entropytemp} \ee
In the zero temperature limit $T\to 0$, this
reduces to the previous result (\ref{simple}).
On the other hand, in the high temperature
 limit $T\to \infty$, it approaches
\be S_A\simeq \f{\pi c}{3}lT. \label{thermalth} \ee This
 is the same as the thermal entropy for the subsystem $A$
as expected.

\subsection{Massive Theories}
\label{massive thoery} \hspace{5mm} When we are away from a critical
point, the logarithmic scaling law Eq.\ (\ref{simple}) does not
persist for $l>\xi$, where $\xi$ is the correlation length (inverse
of the mass gap). For large $l$ ($\gg\xi$), the entanglement entropy
saturates to a finite value \cite{Vidal03,Cardy} \be
S_A=\mathcal{A}\cdot\f{c}{6}\log\f{\xi}{a} \label{massivee}, \ee
where $\mathcal{A}$ is the number of boundary points that separate
$A$ from its complement. Thus, unlike critical (1+1)D systems, the
area law holds for the massive case. This behavior was studied in
details in several 1D quantum spin chains
\cite{Vidal03,Cardy,Its05,Peschel04}, and QFTs
\cite{Cardy,Casini05a,Casini05b}. In \cite{Cardy}, the result
(\ref{massivee}) is derived from an argument similar to
Zamolodchikov's $c$-theorem. We will mention this proof briefly in
section \ref{EE and central charge}.

\section{Entanglement Entropy in Higher Dimensional CFTs}
\label{(d+1)D CFT}
\setcounter{equation}{0} \hspace{5mm} Now we would like to move on
to the computations of entanglement entropy in higher dimensional
conformal field theories CFT$_{d+1\geq 3}$. This was initiated in
\cite{Bombelli,Srednicki} and a partial list of later results can be
found in
\cite{Kabat,Eisert,Cardy,Wolf05,Gioev05,Casini05b,Das,Barthel06,Fursaev}.
In spite of many progresses, the calculation of
 the entropy is too complicated
to find exact results. This is one motivation
 to consider the holographic  way of
computing the quantity as we will discuss later.

As in the 2D CFT case explained in section \ref{2D CFT},
we assume the
CFT$_{d+1}$ is defined on the $d+1$ dimensional manifold
$\mathbb{R}\times N$. We define the subsystem $A$ as the submanifold
of $N$ at a fixed time $t=t_0\in \mathbb{R}$. The strategy of
calculating the entanglement entropy $S_A$ is the same as in the 2D
case. First find  the reduced trace
$\mathrm{tr}_{A}\rho_A^n$
and then plug this in
(\ref{deri}) to obtain $S_A$. We can compute
$\mathrm{tr}_{A}\rho_A^n$
from the partition function
$Z_n$ on the $n$-sheeted $d+1$ dimensional manifold $M_n$
 as in the 2D case (\ref{nsheetp})
\be
\mathrm{tr}_{A} \rho_A^n=\f{Z_n}{(Z_1)^n}.
\label{nsp} \ee
The $n$-sheeted manifold $M_n$ can be constructed as
 follows. First we remove the
infinitely thin $d$ dimensional slice $A$ from
$M_1=\mathbb{R}\times N$. Then the boundary of such a space consists of two $A$s, which we
call $A_{up}$ and $A_{down}$. Next we prepare $n$ copies of such a
manifold. Their boundaries
 are denoted by $A^{i}_{up}$ and $A^i_{down}\ \
  (i=1,2,\ddd,n)$. Now we glue $A^{i}_{up}$ with
 $A^{i+1}_{down}$ for every $i$.
As we take the trace of $\rho_A^n$,
$A^{i=n}_{up}$ is glued with  $A^{1}_{down}$.
In the end this procedure leads to
a 
manifold $M_n$ with conical singularities
where all $n$ cuts meet.

It is not straightforward
 to calculate $Z_n$ for an arbitrary choice of
  $A$ even in free field
theories. This is because the conformal
structure is not as strong as in the 2D CFT case.
Thus below we mainly restrict our arguments
 to specific  forms of $A$ given by the following
two examples. We also simply assume $N=\mathbb{R}^{d}$.

\begin{figure}
\begin{center}
\includegraphics[height=5cm,clip]{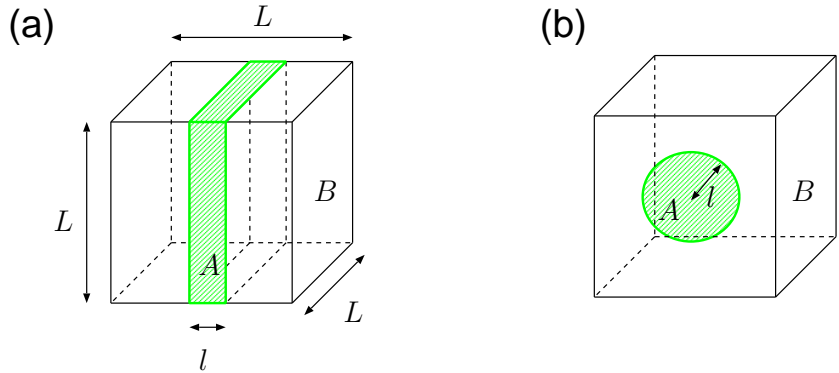}
\end{center}
\caption{
\label{fig: shape_of_A.eps}
Two different shapes of the submanifold $A$
considered in this paper.
(a) The straight belt $A_S$ and (b) the circular disk $A_D$.
(Here, $d=3$ for simplicity.)
}

\end{figure}

The first one is the straight belt of width $l$
\be A_{S}=\{x_i|x_1\in
[-l/2,l/2], x_{2,3,\ddd,d}\in [-\infty,\infty]\}, \label{straighta}
\ee as depicted in Fig.\ \ref{fig: shape_of_A.eps}.
Since the lengths in the directions of $x_2,x_3,\ddd,x_d$ are
infinite, we often put the regularized length $L$.
  Taking the limit $l\to \infty$ and looking at
the region near $x_1=-l/2$, we obtain the subsystem $A_{SL}$  which covers a
half infinite space of $\mathbb{R}^{d}$. The
 boundary in this case is given by  the
straight surface
$\de A_{SL}=\mathbb{R}^{d-1}$.

The second example
is the circular disk $A_{D}$ of
 radius $l$ defined by \be A_{D}=\{x_i|r\leq l\} ,
\label{diska}\ee where $r=\s{\sum_{i=1}^d x_i^2}$ (see Fig.\ \ref{fig:
shape_of_A.eps}.).

\subsection{Entanglement Entropy of $d+1$ D Massless Free Fields}
\label{EE, 1+1 D massive} \hspace{5mm} As an explicit example, we
consider the entanglement entropy of $d+1$ dimensional CFT on
$\mathbb{R}^{1,d}$ defined by massless free fields such as a
massless scalar field or Dirac (or Majorana) fermion. This can be
regarded as the infinite volume limit $L\to \infty$ of the CFT on
 $M=\mathbb{R}^{1,1}\times T^{d-1}$, where the  volume of torus is $L^{d-1}$.

Because this theory is free, we can perform
the dimensional reduction on $T^{d-1}$ and obtain
 infinitely many two dimensional free massive theories
whose masses are given by \be m^2=\sum_{i=2}^{d}k_i^2=
\left(\f{2\pi}{L}\right)^2\cdot \sum_{i=2}^{d}n_i^2,
\label{kkmass}\ee where $k_i=\f{2\pi n_i}{L}$ are the quantized
momenta such that $n_i\in \mathbb{Z}$ in
 the torus directions.

To take this advantage we concentrate on the case where
 the subsystem $A$ is defined by
the straight belt $A_S$ with radius $l$ (\ref{straighta}).
 The point is that the computation of
the  entanglement entropy $S_A$ in this case
is now reduced to the calculations of $S_A$ in
massive 2D QFTs.

\subsubsection{Rough Estimation}
\label{rougth estimate}
\hspace{5mm}
As we reviewed in section \ref{2D CFT},
we know the formulas of
entanglement entropy both in the massless limit (i.e. $l\ll \xi$)
(\ref{simple}) and the massive limit (\ref{massivee}) (i.e.
$l\gg \xi$). The correlation length is estimated as $\xi \sim m^{-1}$,
where $m$ is defined in (\ref{kkmass}). This leads to the following
rough estimation of $S_A$ by replacing the summations of infinitely
many modes $n_i$ with the integral of $k_i$ in the $L\to \infty$
limit
\begin{eqnarray}
S^{rough}_A &=& \sum_{k_2,\cdots,k_d}^{ \xi \le l}
\frac{c}{3}\log \frac{\xi}{a} + \sum_{k_2,\cdots,k_d}^{ \xi \ge l}
\frac{c}{3}\log \frac{l}{a}
\nonumber \\
&=& \left( \frac{L}{2\pi} \right)^{d-1} \frac{c}{3} \left[
\int^{a^{-1}}_{l^{-1}} {d}^{d-1} k \log \frac{\xi}{a} +
\int^{l^{-1}}_{0} {d}^{d-1} k \log \frac{l}{a} \right]
\nonumber \\
&=&\f{c}{3(d-1)\cdot 2^{d-1}\pi^{\f{d-1}{2}} \Gamma(\f{d+1}{2})}
\left[ \frac{L^{d-1}}{a^{d-1}} - \frac{L^{d-1}}{l^{d-1}} \right].
\label{rough}
\end{eqnarray}

If we set $d=3$ (i.e. massless fields in 4 dimension), we obtain \be
S^{rough}_A=\f{c}{24\pi}\left(\f{L^2}{a^2}-\f{L^2}{l^2}\right). \ee
Notice that $c$ is the two dimensional central charge and thus $c=1$
for a 4D real scalar field and $c=1$ (or $c=2$) for a 4D Majorana
(or Dirac) fermion. As can be seen from the exact computation
discussed in the next subsection, this rough estimation already
captures the correct functional form of the entanglement entropy.

The first term in (\ref{rough}) represents the leading divergence
which indeed obeys the area law (\ref{divarea}). This part can be
found by taking the limit $l\to \infty$ i.e. when $A$ is the
straight surface $A_{SL}$. It is also possible to compute this term
analytically as done in \cite{Kabat,Cardy}. On the other hand, the
second term does not depend on the cutoff and thus is an interesting
quantity to examine in more detail.

 The violation of the area law for systems with a finite Fermi
surface can be also understood from this rough estimation of the
entanglement entropy. For simplicity, we assume a spherical Fermi
surface with $k_F$ being the Fermi momentum. For the momentum
$\boldsymbol{k}$ outside and close to the Fermi surface, the gap is
given by $m \sim \xi^{-1}\sim |\boldsymbol{k}|-k_F$. Thus, as
before, the entanglement entropy is estimated as
\begin{eqnarray}
S_A
&=&
\left(
\frac{L}{2\pi}
\right)^{d-1}
\frac{c}{3}
\left[
\int^{a^{-1}}_{|\boldsymbol{k}|=k_F+l^{-1}} d^{d-1} k
\log \frac{\xi}{a}
+
\int^{|\boldsymbol{k}|=k_F+l^{-1}}_{0} d^{d-1} k
\log \frac{l}{a}
\right].
\end{eqnarray}
We thus find, for $l\to \infty$,
\begin{eqnarray}
S_A &\sim& \frac{c}{3}
\frac{2\pi^{\f{d-1}{2}}k_{F}^{d-1}}{(d-1)\Gamma((d-1)/2)} \left(
\frac{L}{2\pi} \right)^{d-1}
 \log \frac{l}{a}
+
\mbox{subleading terms},
\end{eqnarray}
where note that $k_F \propto a^{-1}$. A more precise calculation
based on the Widom conjecture in \cite{Gioev05} gives the
prefactor in front of $L^{d-1}\log l/a$ as the double integral over
the fermi surface in the momentum space and the region $\partial A$
in real space.

\subsubsection{Exact Estimation from Entropic $c$-function}
\label{exact estimate from c-func} \hspace{5mm} The previous
approximation (\ref{rough}) uses the formulas which are exact only
in the two opposite limits $\xi\to \infty$ and $\xi\to 0$. To
perform an exact estimation, we need to be precise about the
intermediate region $\xi\sim l$. In other words, we need to use a
sort of $c$-function under the massive deformation corresponding to
the interpolating region instead of the UV central charge in
(\ref{rough}). To make this more explicit we can employ the entropic
$c$-function $C$ introduced in \cite{Casinicth,Casini05a,Casini05b}.
It is defined for 2D CFTs as follows
\begin{eqnarray}
l \frac{dS_A(l)}{dl} &=& C(lm), \label{entropicf}
\end{eqnarray}
where $l$ is the length of the subsystem $A$ and $m$ is the mass of
the field. For massive free fermions and scalar fields, the function
$C$ is characterized as a solution to a differential equation of
Painleve V type and its numerical form can be found in
\cite{Casini05a,Casini05b}. Unfortunately, its analytical expression
is not known.

\begin{figure}
\begin{center}
\includegraphics[width=8cm,clip]{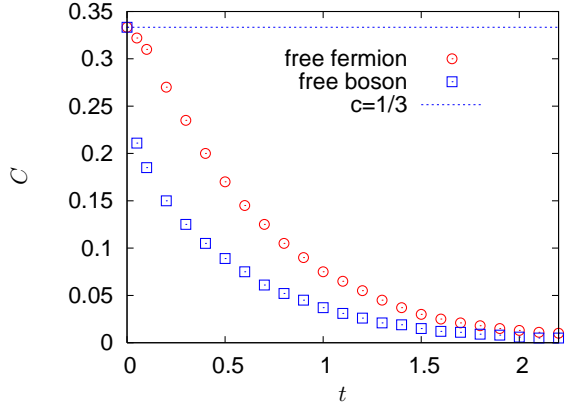}
\end{center}
\caption{
\label{fig: entropy_c_fn.eps}
The entropic $c$-functions $C(x)$ for
free massive real scalar boson and
free Dirac fermion in 1+1 D
reproduced from
\cite{Casini05b}.
}

\end{figure}

This function $C(x)$ is positive and is also a monotonically
decreasing function \cite{Casinicth} with respect to $x$ as in the
Zamolodchikov's $c$-function \cite{Zamo}. These properties are
indeed true in explicit examples \cite{Casini05b}, which we
reproduced in Fig.\ \ref{fig: entropy_c_fn.eps} for a free massive
real scalar boson and free Dirac fermion in 1+1 D. The function
$C(x)$ is normalized such that in the UV limit $x=0$ it is related
to the ordinary central charge via $C(0)=c/3$. Note that if we set
$C=C(0)=c/3$, we recover from this equation the well-known result
(\ref{simple}). We will also show this later independently in
(\ref{entropicd}). It was argued that the
positivity of $C(x)$ is connected to a majorization relation for
local density matrices \cite{Vidal03,Latorre04,Latorre05,Zhou05,Orus05}

In our example of the $d+1$ dimensional free field,
we can reduce it to infinitely many massive fields in two dimensions.
Thus in this case we again just have to sum over the discrete
quantum numbers $n_i$. In the limit $L\to\infty$ we can replace the
sum with an integral
\begin{eqnarray}
l\frac{dS_A(l)}{dl} &=& \left[ \frac{L^{d-1}}{(2\pi)^{d-1}} \right]
\int dk_{2}\cdots dk_{d}~ C(l|k|)
\nonumber \\
&=& \left[
\frac{L^{d-1}}{2^{d-2}\pi^{\f{d-1}{2}}\Gamma\left(\f{d-1}{2}\right)}
\right]\int_{0}^{\infty} dk k^{d-2} C(lk). \label{sumdame}
\end{eqnarray}
The merit of the quantity $C$ instead of $S_A$ itself is that it
does not include UV divergences and thus we can set $a=0$ in $C$.
After the integration of $l$ we find
\begin{eqnarray}
S_A(l) &=& \left[
\frac{L^{d-1}}{2^{d-2}\pi^{\f{d-1}{2}}\Gamma\left(\f{d-1}{2}\right)}
\right]\int_{0}^{\infty} dk k^{d-2}
\int_{a}^{l}\frac{d\ti{l}}{\ti{l}} C(\ti{l}k),
\no
&=& \left(2^{d-1}\pi^{\f{d-1}{2}}~\Gamma\left((d+1)/2\right)
\right)^{-1}\cdot \left[\int_0^\infty dx x^{d-2}C(x)\right]\cdot
\left[\f{L^{d-1}}{a^{d-1}}-\f{L^{d-1}}{l^{d-1}}\right] \nonumber \\
&\equiv& K\left[\f{L^{d-1}}{a^{d-1}}-\f{L^{d-1}}{l^{d-1}}\right].
\label{exactst}
\end{eqnarray}
where we determine the integral constant by requiring that $S_A(l)$
should be vanishing\footnote{It is possible that this requirement is
not absolute, i.e. this choice of the cutoff $a$ may depend on the
theory we consider. Thus only the constant $K$ in front of the
second term (i.e. finite term) in (\ref{exactst}) has a qualitative
meaning.} at $l=a$ since we are cutting off degrees of freedom below
the energy scale $a^{-1}$. It is straightforward to find analogous
formula for the free massive fields. This is given just by replacing
$k$ in (\ref{sumdame}) or (\ref{exactst}) with $\s{k^2+m^2}$.

The second term in (\ref{exactst}) does not depend on the cutoff
$a$. Thus we are interested in its coefficient $K$ which is
proportional to the integral of the function $x^{d-2}C(x)$. In
principle, we can compute it numerically based on the numerical
results of $C(x)$. Indeed by this method the coefficient $K$ was
computed for three dimensional free fields in \cite{Casini05b}. We
extend it to four dimensions which we are interested in later
discussions and present the result as follows
\begin{eqnarray}
K &=& \left\{
\begin{array}{ll}
\displaystyle \frac{1}{\pi} \int_{0}^{\infty} \mathrm{d}t C(t)
\simeq 0.039,
 &  \mbox{for $d+1=3$ dimensional real scalar boson} \\
\displaystyle \frac{1}{\pi} \int_{0}^{\infty} \mathrm{d}t C(t)
\simeq 0.072,
&  \mbox{for $d+1=3$ dimensional Dirac fermion} \\
& \\
\displaystyle \frac{1}{4 \pi} \int_{0}^{\infty} \mathrm{d}t t C(t)
\simeq 0.0049,
&  \mbox{for $d+1=4$ dimensional real scalar boson} \\
\displaystyle \frac{1}{4 \pi} \int_{0}^{\infty} \mathrm{d}t t C(t)
\simeq 0.0097,
&  \mbox{for $d+1=4$ dimensional Majorana fermion}. \label{numericalen}\\
\end{array}
\right. \label{eq: k for boson and fermion for generic D no approx}
\end{eqnarray}
To find the coefficient $K$ in higher dimensions it is useful to
notice that when $x$ is large the entropic $c$-function $C(x)$
behaves as ($K_\nu(x)$ is the deformed Bessel function)
\begin{eqnarray}
C_{scalar}(x) \simeq  \frac{1}{4} x K_1 (2x), \ \ \ \mbox{and}\ \ \
C_{Dirac}(x) \simeq  \frac{1}{2} x K_1 (2x), \label{largex}
\end{eqnarray}
for a 2D free scalar field and a 2D Dirac fermion. When the
dimension $d$ is large, the contribution of the integral $\int dx
x^{d-2}C(x)$ mainly comes from the large $x$ region. Thus $K$ can be
well approximated by plugging (\ref{largex}) into (\ref{exactst}).
This leads to\footnote{For example, from this approximation we find
$K_{boson}=0.0497$ and $K_{fermion}=0.00995$ for $d=3$ and these are
rather close to the previous results in (\ref{numericalen}). It may
also be interesting to compare this with the our rough estimation
done in the previous subsection. There we found $K^{rough}_{scalar}=
K^{rough}_{fermion}=\f{1}{24\pi}=0.0133$ when $d=3$.}
 \be
K_{scalar}\simeq
2^{-d-3}\pi^{(1-d)/2}\Gamma\left(\f{d-1}{2}\right),\ \ \
K_{fermion}\simeq 2K_{scalar},
\ee where $K_{scalar}$ corresponds to
a $d+1$ dimensional real scalar field while $K_{fermion}$
to the $d+1$ dimensional fermions which is reduced to a
2D Dirac fermion\footnote{In higher dimension we need to multiply an
appropriate degeneracy with $K_{fermion}$ to obtain the result for a
ordinary fermion such as Dirac fermion.}.

Finally we would like to stress again that our result
(\ref{exactst}) was obtained by assuming a free field theory. In the
presence of interactions we no longer have the simple sum over
infinitely many massive fields in two dimensions (\ref{sumdame}) due
to interactions between two different massive fields.

\subsubsection{Entanglement Entropy of 4D Gauge Field}
\label{EE of 4D gauge field}
\hspace{5mm} As we will consider 4D gauge theories later, we would
also like to examine the entanglement entropy of 4D gauge field. We
neglect the interactions as before and thus we can concentrate on
the abelian gauge theory. Its gauge fixed action with the ghost $c$
is given by
\begin{eqnarray}
S &=& \int d^4 x\left[ \frac{1}{4} F_{\mu\nu}F^{\mu\nu} + \frac{1}{2
\alpha} \left(
\partial_{\mu} A^{\mu}
\right)^2 + \bar c \left(-\Box \right) c \right].
\end{eqnarray}
In order to compute the entanglement entropy, we consider the gauge
theory on an $n$-sheeted manifold $M_n$ as before. We can rewrite
the gauge field action as follows (we fix the gauge by setting
$\ap=1$)
\begin{eqnarray}
S_{A_\mu}&=& \int d^4 x\left[ \frac{1}{4} F_{\mu\nu}F^{\mu\nu} +
\frac{1}{2} \left(
\partial_{\mu} A^{\mu}
\right)^2 \right]
\nonumber \\
&=& \int d^4 x\frac{1}{2}
\partial_{\mu}A^{\nu}\partial^{\mu}A_{\nu}
+ \left[ \int d^3 x [-(\partial_{\nu}A^{\nu})A^{0}
+(\partial_{\mu}A^{0})A^{\mu}] \right]_{surf},
\end{eqnarray}
where $[\ddd]_{surf}$ denotes the surface term when we performed an
partial integration. If we neglect the surface term, the theory is
equivalent to four real scalar fields and a ghost field which is a
complex scalar. Since the complex ghost scalar field cancels two
real scalars, the theory is equivalent to two real scalar fields.
 However, there is a subtle issue on
the surface term that appears when we do the partial integration.
Since fields are discontinuous along the time direction in some
region of the spacetime, we got surface contributions. In this paper
we assume such a term is not relevant for the computation of the
entanglement entropy as\footnote{Indeed if we include such a
contribution we find the total entropy of the gauge field becomes
negative in the particular case discussed in \cite{Kabat}, which
looks strange if we remember the original definition (\ref{eq:def
entanglement entropy}).} in \cite{Kabat}.

\subsection{Entanglement Entropy and Central Charges in 4D CFT}
\label{EE and central charge, 4D} \hspace{5mm} As we have seen, the
entanglement entropy in 2D CFTs is proportional to the central
charge $c$. Since the central charge roughly measures the number of
degrees of freedom $N_{dof}$, we find the entanglement entropy is
also proportional to $N_{dof}$. This fact is very natural as its
name of `entropy' shows. Therefore we may expect that a similar
story is true also in the higher dimensional theories. As such an
example, below we consider 4D CFTs. Indeed we will find that an
important part of the entanglement entropy is proportional to the
central charges. See also \cite{Solodukhinc,Fursaev} for an earlier
discussion.

In principle, it is possible to extend the relation between central
charges and entanglement entropy to higher dimensions as far as the
spacetime dimension is even. When we consider odd dimensional
spacetime, we do not have any clear definition of central charges
due to the absence of the Weyl anomaly. Under this
situation,
the entanglement entropy may play an
important alternative role
\footnote{We are grateful to Anton Kapustin for pointing
out this possibility to us.}.


\subsubsection{Entanglement Entropy from Weyl Anomaly}
\label{EE from Weyl anomaly} \hspace{5mm} Central charges in CFTs
can be defined from the Weyl anomaly (or conformal anomaly) $\la
T^\mu_\mu\lb$. Define the energy-momentum tensor $T^{\mu\nu}$ in
terms of the functional derivative of the (quantum corrected) action
$S$ with respect to the metric $g_{\mu\nu}$ \be
T^{\mu\nu}=\f{4\pi}{\s{g}}\f{\delta S}{\delta g_{\mu\nu}}.
\label{emtensor}\ee In 2D CFTs, the Weyl anomaly is given by the
well-known formula \be \la T^\mu_{\mu}\lb =-\f{c}{12}R,
\label{tdanomaly}\ee where $R$ is the scalar curvature. We can
regard this as a definition of the central charge $c$ in 2D CFTs.

Now we move on to 4D CFTs. In our normalization of
(\ref{emtensor}), the Weyl anomaly can be written as \be \la
T^\ap_\ap\lb=-\f{c}{8\pi}W_{\mu\nu\rho\sigma}W^{\mu\nu\rho\sigma}
+\f{a}{8\pi}\ti{R}_{\mu\nu\rho\sigma}\ti{R}^{\mu\nu\rho\sigma}.
\label{catracer} \ee in a curved metric background $g_{\mu\nu}$,
where $W$ and $\ti{R}$ are the Weyl tensor and the dual of the
curvature tensor. Notice that the second term is the Euler density.
In terms of the ordinary curvature tensor, we can express the
curvature square terms in (\ref{catracer}) as follows \ba
W_{\mu\nu\rho\sigma}W^{\mu\nu\rho\sigma}&=&
R_{\mu\nu\rho\sigma}R^{\mu\nu\rho\sigma}
-2R_{\mu\nu}R^{\mu\nu}+\f{1}{3}R^2, \no
\ti{R}_{\mu\nu\rho\sigma}\ti{R}^{\mu\nu\rho\sigma}&=&
R_{\mu\nu\rho\sigma}R^{\mu\nu\rho\sigma} -4R_{\mu\nu}R^{\mu\nu}
+R^2. \ea

The coefficients $c$ and $a$ in (\ref{catracer}) are
called\footnote{ The central charge $a$ should not be confused with
a UV cutoff. To avoid confusion, $a_{cutoff}$ is used to denote the
UV cutoff in this subsection. } the central charges of 4D CFTs
\cite{Cardya,Osborn,AFGJ}. This is the original definition of the
central charges in 4D CFTs. The central charge $a$ is believed to
decrease monotonically under the renormalization group (RG) flow,
while for $c$ this is not true and indeed counter examples are
known; these properties of the central charges $a$ and $c$ are
confirmed in many supersymmetric examples e.g.\ \cite{AFGJ}.

To compute the entanglement entropy, we first consider the partition
function $Z_{n}$ on the $d+1$ dimensional
$n$-sheeted manifold
$M_n$. Then we find the trace of $\rho^n$
 reduced to the subsystem $A$ is given by the formula (\ref{nsp}).
 The entanglement
 entropy can be found by taking the derivative of $n$ with
the $n\to 1$ limit. If we define the length scale of the manifold
$A$ by $l$,
 then the scaling of $l$ is related to the Weyl scaling. They should
 be the same\footnote{Equally we can say that the scaling of $l$ is
 oppositely related to the scaling of the cutoff i.e. $l\f{d}{dl}
 =-a_{cutoff}\cdot\f{d}{da_{cutoff}}$}
 at least in the $n\to 1$ limit. In this way we find \ba
l\f{d}{dl}\log
[\mathrm{tr}_{A}\,\rho_A^n] &=&2\int d^{d+1}x\
g_{\mu\nu}(x)\f{\delta}{\delta g_{\mu\nu}(x)}
\left[
\log Z_n
- n \log Z_1
\right]
\no
&=&
-\f{1}{2\pi}\left\la \int d^{d+1}x~ \s{g}T^\mu_\mu(x)
\right\lb_{M_n}+\f{n}{2\pi}\left\la \int d^{d+1}x~ \s{g}T^\mu_\mu(x)
\right\lb_{M_1}. \ea When we consider a CFT on $M=\mathbb{R}^{d+1}$,
the second term (i.e. integral on $M_1=M=\mathbb{R}^{d+1}$) become
obviously vanishing. Below we omit writing the second term
explicitly just to make the appearance of expressions simple even if
$M$ is a curved manifold. Then the entanglement entropy satisfies
\ba l\f{d}{dl}S_A &=&-\lim_{n\to 1}l\f{d}{dl}\left(\f{\de}{\de
n}\log [\mathrm{tr}_A\,\rho_A^n]\right) \no &=&
\f{1}{2\pi}\lim_{n\to 1} \f{\de}{\de n}\left\la \int d^{d+1}x~
\s{g}T^\mu_\mu(x) \right\lb_{M_n}. \label{tdent}\ea

\subsubsection{Entanglement Entropy and Central Charges}
\label{EE and central charge} \hspace{5mm} Eq.\ (\ref{tdent}) can be
used to relate the entanglement entropy and central charge in a
direct fashion. Let us apply (\ref{tdent}) to 2D CFTs first. We
assume the submanifold $A$ is a segment of the length $l$ in the
total system. Then, the $n$-sheeted manifold $M_n$ has two conical
singularities at $u$ and $v$ that separate $A$ and $B$. If one goes
around these singularities, one picks up $2\pi n$ phase, i.e.,
$2\pi(n-1)$ extra phase compared with $2\pi$. (See Fig.\ \ref{fig:
n-riemann.eps}.) These singularities are reflected in the Euler
number \be \chi[M_n]=\f{1}{4\pi}\int_{M_n} d^2 x\s{g}R=2(1-n), \ee
where we noted the scalar curvature is given by $R=4\pi
(1-n)[\delta^{(2)}(u)+\delta^{(2)}(v) ]$ in the presence of a
deficit angle $2\pi(1-n)$ at the conical singularities. Plugging
(\ref{tdanomaly}) into (\ref{tdent}), we obtain \be
l\f{d}{dl}S_A=-\f{\de}{\de n}\left(\f{c}{24\pi}\int
d^2x\s{g}R\right)=\f{c}{3} \label{entropicd}. \ee We thus reproduce
the known result (\ref{simple}) (see also (\ref{entropicf})).

It is also possible to derive (\ref{massivee}) from (\ref{tdent}) by
noting that
\begin{eqnarray}
m\frac{\partial S_A}{\partial m} &=& l\frac{\partial S_A}{\partial
l} = \frac{1}{2\pi} \lim_{n\to 1} \frac{\de}{\de n} \left\la \int
d^{2}x~ \s{g} T^{\mu}_\mu \right\rangle.
\end{eqnarray}
 When $A=A_{SL}$ (i.e., $\mathcal{A}=1$ in (\ref{massivee}) ), the
integral on the right hand side is evaluated as
\begin{eqnarray}
\int d^{2}x~ \s{g} \left\la T^\mu_\mu \right\rangle &=& - \pi
\frac{c}{6} \left( n-\frac{1}{n} \right),
\end{eqnarray}
by an argument similar to Zamolodchikov's $c$-theorem
\cite{Cardy}. We thus recover (\ref{massivee}).

If we repeat the same analysis in 4D CFTs, we find
\begin{eqnarray}
l\f{d}{dl}S_A
\!\!\! &\! =\! & \!\!\!
\lim_{n\to 1} \f{\de}{\de n}
\left[-
\f{c}{16\pi^2}\!\! \int_{M_n} \!\!d^4x \s{g}
W_{\mu\nu\rho\sigma}W^{\mu\nu\rho\sigma}
+
\f{a}{16\pi^2}\int_{M_n}\!\! d^4 x\s{g}
\ti{R}_{\mu\nu\rho\sigma}\ti{R}^{\mu\nu\rho\sigma}
\right]\label{entfin}\\
&=&
\gamma_1\cdot \f{{\mbox{Area}}(\de A)}{a^2_{cutoff}}+\gamma_2,
\label{entfinr}
\end{eqnarray}
where $\gamma_1$ and $\gamma_2$ are numerical
constants. The first term in (\ref{entfinr}) comes from the integral
of the $W^2$ term in (\ref{entfin}) and represents the leading
divergence $\sim a_{cut off}^{-2}$. This is because the curvature
tensor is divergent as $R\sim a_{cut off}^{-2}$ at the surface $\de
A$, where the deficit angle presents and behaves like a delta
function supported on $\de A$. The Euler density term does not have
such a divergence since it is a topological invariant. Thus the
constant $\gamma_1$ is proportional to $c$. Another constant
$\gamma_2$ comes from both terms in
 (\ref{entfin}) and it is proportional to the linear combination of
 $a$ and $c$. By integrating (\ref{entfinr}), we can express the
 entanglement entropy as follows
\be S_A=\f{\gamma_1}{2}\cdot\f{{\mbox{Area}}(\de A)}{a^2_{cutoff}}
+\gamma_2\log\f{l}{a_{cutoff}}+S^{others}_A, \label{fouren}\ee where
the final term $S^{others}_A$ expresses terms which are independent
of the total scaling $l\to e^{\ap}l$. In other words, $S^{others}_A$
depends on the detailed shape of the surface $\de A$. In this way,
the central charges determine the entanglement entropy up to these
contributions $S^{others}_A$. Notice that the leading divergence
(\ref{fouren}) agrees with the area law (\ref{divarea}). In our
later arguments using AdS/CFT duality, the gravity computations in
section \ref{AdS d+2/CFT d+1} reproduce the same behavior as
(\ref{fouren}). When we assume $a=c$, both $\gamma_1$ and $\gamma_2$
are proportional to $a$. This also agrees with our later gravity
computations in section \ref{EE 4D CFT}. For example, in the
${\mathcal{N}}=4$ $SU(N)$ super Yang-Mills the central charges are
given by $a=c=(N^2-1)/4$ \cite{Gubser}\ and thus they satisfy the
condition.

In particular, when $A$ is the circular disk $A_D$ with radius $l$,
 the system only depends on $l$ and
$a_{cutoff}$. Thus the trace anomaly completely determines the
entanglement entropy $S_A$. On the other hand, in the case of the
straight belt $A_S$ there are two length
 scales $l$ and $L$ and the result (\ref{fouren}) becomes less
 predictive. Indeed the finite term which we discussed before
takes the form $\propto \f{L^2}{l^2}$ and thus it is included in
$S^{others}_A$ in (\ref{fouren}).  Since this term is not directly
related to the central charges, we expect that its value may be
shifted when we change the t' Hooft coupling as is so in the thermal
entropy. Indeed, the comparison of the numerical results from the
AdS$_5\times$ S$^5$ and the free ${\mathcal N}=4$ super Yang-Mills
supports this speculation as we will see in section
\ref{EE 4D CFT}.

Even though the constant $\gamma_1$ is not universal in the sense
that it depends on the choice of the UV cut off, the other one
$\gamma_2$ is universal and an interesting quantity to evaluate. In
principle, this is reduced to a differential geometric computation.
Since the evaluation of total expression turns out to be rather
complicated, below we would like to compute some particular
important terms.

It is straightforward to evaluate the contribution from the second
term (Euler density) in (\ref{entfin}) because this is a topological
term. As shown in \cite{FursaevR}, in a 4D manifold $M_{n}$ with a
codimension two surface $\Sigma$ around which conical singularities
develop (with a deficit angle $2\pi (1-n)$) we obtain \be
\chi[M_n]=\f{1}{32\pi^2}\int_{M_{n}}
d^4 x\s{g}\ti{R}\ti{R}=(1-n)\chi[\Sigma]
+\f{1}{32\pi^2}\int_{M_n-\Sigma}d^4 x \s{g}\ti{R}\ti{R}, \ee where
$M_n-\Sigma$ denotes the smooth manifold defined by subtracting the
singular part $\Sigma$ from $M_n$. Therefore the contribution of the
$\ti{R}^2$ term in (\ref{entfin}) to the constant $\gamma_2$ is
given by \be \gamma_2^{top}=-2a\cdot \chi[\de A],\ \ \ \ \
 (\mbox{especially},\ \
 \gamma^{top}_2=-4a\ \mbox{when}\ \ \de A={\mbox{ S}}^2). \label{topt}\ee

To make the analysis of the $W^2$ term in (\ref{entfin}) simple,
below we only consider the case where the second fundamental form
(or the extrinstic curvature) of $\de A$, when embedded in the 4D
manifold $M_n$, can be neglected. This is true when we consider the
straight belt $A_S$. Another typical example is when $M_n$ is an
Euclidean black hole and $\de A$ is its horizon. We also concentrate
on the case where $\de A$ is a connected manifold. Under these
assumptions we can employ the differential geometric results in
\cite{FursaevR}
\begin{eqnarray}
&&\int_{M_{n}}\!
d^4 x\s{g}R^2\!-\!\int_{M_{n}-\Sigma}\!\!\!\!\!\!\!\!
d^4x\s{g}R^2
=8\pi(1-n)\int
(R_{\Sigma}+2R_{ii}-R_{ijij})+{\mathcal{O}}((1-n)^2),\no &&
\int_{M_{n}}\! d^4 x\s{g}R^{\mu\nu\rho\sigma}R_{\mu\nu\rho\sigma}\!-
\!\int_{M_{n}-\Sigma}\!\!\!\!\!\!\!\!
d^4 x\s{g}R^{\mu\nu\rho\sigma}R_{\mu\nu\rho\sigma} =8\pi(1-n)\int
R_{ijij}+{\mathcal{O}}((1-n)^2),
\no && \int_{M_{n}}\!
d^4 x\s{g}R^{\mu\nu}R_{\mu\nu}\!-\!\int_{M_{n}-\Sigma}\!\!\!\!\!\!\!\!
d^4 x\s{g}R^{\mu\nu}R_{\mu\nu} =4\pi(1-n)\int
R_{ii}+{\mathcal{O}}((1-n)^2),
\end{eqnarray}
where $R_{\Sigma}$ is the intrinsic curvature of the 2D submanifold
$\Sigma$; $R_{ij}$ and $R_{ijkl}$ denote the curvature tensors
projected onto the direction normal to $\Sigma$ (e.g.
$R_{ij}=R_{\mu\nu}n^{\mu}_in^{\nu}_j$ using the two orthonormal
vectors $n^i_{\mu}\ \ \ (i=1,2)$ orthogonal to $\Sigma$). In the
end, we obtain\footnote{Refer also to \cite{Solodukhin} for an
earlier computation of a similar expression of the logarithmic term
from a different approach.}
 (this includes both contributions from
$W^2$ and $\ti{R}^2$)
\begin{eqnarray}
&&\gamma_2= \f{c}{6\pi}\int_{\Sigma=\de A}d^2
x\s{g}\left(R_{\Sigma=\de A}+2R_{ijij}-R_{ii}\right)
-\f{a}{2\pi}\int_{\Sigma=\de A}d^2 x\s{g}R_{\Sigma=\de A}.
\end{eqnarray}
Especially when  $a=c$,
\begin{eqnarray}
\gamma_2= -\f{a}{6\pi}\int_{\Sigma=\de A}\!\!\!\! d^2
x\s{g}\left(2R_{\Sigma=\de A}-2R_{ijij}+R_{ii}\right). \label{aceq}
\end{eqnarray}
 under the previous assumption that the second
fundamental form is zero. We will later compare this result with the
one from gravity side in section \ref{EE 4D CFT}.

\section{Holographic Interpretation}
\label{holographic} \setcounter{equation}{0} \hspace{5mm} The main
purpose of this paper is to compute the entanglement entropy in
$d+1$ dimensional conformal field theories CFT$_{d+1}$ via the
AdS/CFT correspondence. This duality relates the
 CFT$_{d+1}$ to the $d+2$ dimensional AdS space AdS$_{d+2}$. Then we
 expect that the entanglement entropy can be
 computed as a geometrical quantity
 in the AdS$_{d+2}$ space just as the thermal entropy of CFTs is found from the
 area formula of AdS black hole entropy \cite{GKP}.

As in section \ref{2D CFT} the CFT$_{d+1}$ is defined on $M={\mathbb{R}}\times
N$ and we divide $N$ into two regions $A$ and $B$. We assume the
space-like $d$ dimensional manifold $N$ is now given by ${\mathbb{
R}}^d$ or S$^d$ such that $M$ is the boundary of AdS$_{d+1}$ in the
Poincare coordinates
\be ds^2=R^2~\f{dz^2-dx_0^2+
\sum_{i=1}^{d-1}dx_i^2}{z^2},
\label{Poincare} \ee
or the global
coordinates \be  ds^2=R^2\left(-\cosh^2\rho
dt^2+d\rho^2+\sinh\rho^2d\Omega_{d}^2\right), \label{global}\ee
respectively.

\subsection{General Proposal}
\label{general proposal} \hspace{5mm} In this setup we propose that
the entanglement entropy $S_A$ in CFT$_{d+1}$ can be computed from
the following area law relation
\begin{equation} S_{A}=\frac{{\rm Area}(\gamma_{A})}{4G^{(d+2)}_N}.
\label{arealaw}
\end{equation}
The manifold $\gamma_{A}$ is the $d$-dimensional static minimal
surface in AdS$_{d+2}$ whose boundary is given by $\de A$. Its area
is denoted by ${\rm Area}(\gamma_{A})$. Also $G^{(d+2)}_{N}$ is the
$d+2$ dimensional Newton constant. It is obvious that the leading
divergence $\sim a^{-(d-1)}$ in (\ref{arealaw}) is proportional to
the area of the boundary $\de A$ and this agrees with the known
property (\ref{divarea}).

This proposal is motivated by the following physical interpretation.
Since the entanglement entropy $S_A$ is defined by smearing out the
region $B$, the entropy is considered to be the one for an observer
in $A$ who is not accessible to $B$. The smearing process produces
the fuzziness for the observer and that should be measured\footnote{
It may be interesting to note that this origin of entropy is
somewhat analogous to the recently proposed `fuzzball' picture (for
a review see \cite{fuzzy}).} by $S_A$. In the higher dimensional
perspective of the AdS space, such an fussiness appears by hiding a
part of the bulk space AdS$_{d+2}$ inside an imaginary horizon,
which we call $\gamma$.  It is clear that $\gamma$ covers the
smeared region $B$ from the inside of the AdS space and thus we find
$\de \gamma=\de B(=\de A)$. We expect that it is the holographic
screen for the hidden part in the bulk. To choose the minimal
surface as in (\ref{arealaw}) means that we are seeking the severest
entropy bound \cite{holography,BiSu,Bousso} for the lost
information. In the examples of AdS$_3/$CFT$_2$, we will show below
that the bound is actually saturated. Therefore it is natural to
expect that the bound is always saturated even in the higher
dimensional $(d\geq 2)$ cases. These considerations lead to our
proposal (\ref{arealaw}). Notice also that the properties
(\ref{ext}) and (\ref{extt}) are obviously satisfied for
(\ref{arealaw}).

It is also straightforward to extend this formula (\ref{arealaw}) to
any asymptotically AdS spaces and we argue that the claim remains
the same in these generalized cases. For example, if we consider a
AdS Schwarzschild black hole, then the minimal surface $\gamma_{A}$
wraps the part of its real horizon
 as we will see later in section \ref{finite temperature case}.
This consideration fixes
 the normalization of (\ref{arealaw}).

\subsection{Intuitive Derivation from AdS/CFT}
\label{intuitive deriv} \hspace{5mm} Let us try to understand how
the area law (\ref{arealaw}) can be derived from known facts on
AdS/CFT correspondence. As we have seen in section \ref{2D CFT}, it
is essential to compute $\mathrm{tr}_A\,\rho_A^n$ in order to obtain
the entanglement entropy. It is equivalent to the partition function
of the CFT on the multiple (i.e. $n$ times) covered space. Then
$S_A$ can be found from the formula (\ref{deri}).

Let us start with the AdS$_3/$CFT$_2$ example with a single
interval. In this case as we have seen, $\mathrm{tr}_A\,\rho_A^n$ is
equivalent to the $n$ products of the two point functions $\la
\Phi^{+(k)}_{n}\Phi^{-(k)}_n\lb$ as in (\ref{densityend}). The
conformal dimension of $\Phi^{(k)\pm}_{n}$ is given by
$\Delta_{n}=\f{c}{24}(1-n^{-2})$. The CFTs on disconnected $n$
sheets (remember the description explained in section \ref{2D CFT})
is equivalent to a CFT on a single sheet ${\mathbb{R}}^2$ whose
central charge is $nc$ with two twisted vertex operators
$\Phi^+_{n}$ and $\Phi^-_n$ (distinguish them from
$\Phi^{(k)\pm}_{n}$) inserted.

 In AdS/CFT\footnote{Here we consider the AdS dual of the CFT with central
 charge $nc$. Finally we take the limit $n\to 1$.},
 such a two point function $\la
\Phi^+_{n}(P)\Phi^-_n(Q)\lb$ in the CFT can be computed as \be \la
\Phi^+_n (P)\Phi^-_n(Q)\lb \sim e^{-\f{2n\Delta_{n}\cdot
L_{PQ}}{R}}, \ee where $L_{PQ}$ is the geodesic distance between $P$
and $Q$. Therefore we can derive explicitly the area law
(\ref{arealaw}) as follows \be S_{A}=2\left( \left. \f{\de
(n\Delta_n)}{\de n}\right|_{n=1}\right)\cdot
\f{L_{\gamma_A}}{R}=\f{L_{\gamma_A}}{4G^{(3)}_N},\ee from AdS/CFT
correspondence.

In higher dimensions, we can again compute $\mathrm{tr}_A\,\rho_A^n$
as the path-integral over the multi covered space with $n$ sheets.
We expect that this system is equivalent to a CFT on
$\mathbb{R}^{1,d}$ which has the $n$ replica fields $\phi_i(x^\mu)$
with a twist-like operator $\Phi_n$ inserted (assuming the simplest
case that $\de A$ is a connected manifold). Notice that this
operator is localized in codimension two subspace of
$\mathbb{R}^{1,d}$. Then $\mathrm{tr}_A\,\rho_A^n$ is equal to the
one point function $\la \Phi_n\lb$. As in the Wilson loop operator
case \cite{Wilsonline}, we naturally expect that it can be computed
as \be \la \Phi_n\lb\sim e^{-\ap_n\mbox{Area}_A}, \label{dgeomin}
\ee where Area$_A$ is the area of the minimal surface in AdS$_{d+2}$
whose boundary is $\de A$; $\ap_n$ is a $n$-dependent constant
($\lim_{n\to\infty}\f{\ap_n}{n}=$finite).

This form (\ref{dgeomin}) is almost clear from the following
argument. First we notice that $\log \la \Phi_n\lb$ should be equal
to the factor $\f{1}{G^{(d+2)}_N}$ times a certain diffeomorphism
invariant quantity as is clear from the supergravity side. Then the
latter should have the momentum dimension $-d$. Only such a
candidate is essentially the area term as in (\ref{dgeomin}),
assuming that it is given by a local integral.

Applying the formula (\ref{deri}) we find \be S_A= \left( \left.
\f{\de \ap_n}{\de n}\right|_{n=1}\right)\cdot \mbox{Area}_A. \ee The
coefficient can be fixed by requiring that the entanglement entropy
at a finite temperature should be reduced to the thermal entropy
(i.e. black hole entropy) when $A$ is the total space (see also
later discussions in section 7 on this point). This leads to
(\ref{arealaw}).

\section{Entanglement Entropy in 2D CFT from AdS$_3$}
\label{AdS3/CFT2} \setcounter{equation}{0} \hspace{5mm} We start
with the AdS$_3$ $(d=1)$ in the global coordinates (\ref{global}).
According to AdS/CFT correspondence \cite{Maldacena}, the
gravitational theories on this space are dual to $1+1$ dimensional
conformal field theories with the central charge \cite{BH} \be
c=\f{3R}{2G^{(3)}_N},\label{centralads}\ee where $G^{(3)}_N$ is the
Newton constant in three dimensional gravity\footnote{ Remember that
$G^{(d+2)}_N$ is defined such as $S_{gravity}=\f{1}{16\pi
G^{(d+2)}_N}\int d^{d+2}x \s{g}R+....$ for any dimension $d$.}.

\subsection{AdS$_3$ Space and UV Cutoff in Dual CFTs}
\label{AdS3, cutoff}
\hspace{5mm} At the boundary $\rho=\infty$ of the AdS$_3$, the
metric is divergent. To regulate relevant physical quantities we
need to put a cutoff $\rho_0$ and restrict the space to the bounded
region $\rho\leq\rho_0$. This procedure corresponds to the ultra
violet (UV)  cutoff in the dual conformal field theory
\cite{ADSWitten,SuWi}. If we define the dimensionless UV cutoff
$\delta$ ($\propto\ length$), then we find the relation
$e^{\rho_0}\sim \delta^{-1}$. In the example of the previous
section, $\delta$ should be identified with \be e^{\rho_0}\sim
\delta^{-1}=L/a.\label{delti} \ee Remember that $L$ is the total
length of the system and $a$ is the lattice spacing (or UV cutoff).
Notice that there is actually an ambiguity about the ${\mathcal
O}(1)$ numerical coefficient in this relation\footnote{However, this
ambiguity does not affect universal quantities which do not depend
on the cut off $a$ and we will consider such quantities in the later
arguments.}.

The holographic principle tells us that true physical degrees of
freedom of the gravitational theory in some region is represented by
its boundary of that region. This is well-known in the black hole
geometries and it leads to the celebrated area law of the
Bekenstein-Hawking entropy. In the context of AdS/CFT correspondence
degrees of freedom in AdS$_{d+1}$ space are represented by its
 boundary of the form ${\mathbb{R}}_{t}\times$ S$^{d-1}$, where the
 dual conformal field theory lives. We can compute the number of
 degrees of freedom $N_{dof}$ by applying
 the area law in three dimensional spacetimes
 to the boundary in the AdS$_3$ space \cite{SuWi} . This
 leads to the following estimation
 \be N_{dof}\sim  \f{{\rm Boundary\ Length}}{4G^{(3)}_N}
 = \f{2\pi R\sinh\rho_0}{4G^{(3)}_N}
 \simeq \f{\pi c}{6}\cdot \f{L}{a}. \label{degfr} \ee The central
 charge $c$ is roughly proportional to the number of fields. The ratio
 $L/a$ counts the number of independent points in the presence of
 the lattice spacing $a$. Therefore the result (\ref{degfr})\ agrees with what we expect
 from the conformal field theory at least up to the unknown numerical
 coefficient.

\subsection{Geodesics in AdS$_3$ and Entanglement Entropy in CFT$_2$}
\label{geodesics in AdS3} 
\begin{figure}
\begin{center}
\includegraphics[height=4cm,clip]{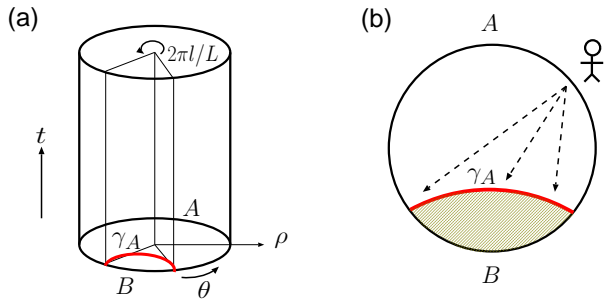}
\end{center}
\caption{
\label{fig: ads3_cft2.eps}
(a)
AdS$_3$ space and CFT$_2$ living on its boundary
and (b) a geodesics $\gamma_A$ as a
holographic screen.
}
\end{figure}

\hspace{5mm} In the global coordinate of
AdS$_3$ (\ref{global}), the $1+1$ dimensional spacetime, in which
the CFT$_2$ is defined, is identified with the cylinder
$(t,\theta(\equiv \Omega_1))$ at the (regularized) boundary
$\rho=\rho_0$. Then we consider the AdS dual of the setup in section
\ref{deriv EE on circ}. The subsystem $A$ corresponds to $0\leq
\theta\leq 2\pi l/L$ and we can discuss the entanglement entropy by
applying our proposal (\ref{arealaw}). In this lowest dimensional
example, the minimal surface $\gamma_{A}$, which plays the role of
the holographic screen \cite{holography,BiSu,Bousso}, becomes one
dimensional. In other words, it is the geodesic line which connects
the two boundary points at $\theta=0$ and $\theta=2\pi l/L$ with $t$
fixed (see Fig.\ \ref{fig: ads3_cft2.eps}) .

Then to find the entropy we calculate the length of  the geodesic
line $\gamma_{A}$. The geodesics in AdS$_{d+2}$ spaces are given by
the intersections of two dimensional hyperplanes
 and the AdS$_{d+2}$ in the ambient $\mathbb{R}^{2,d+1}$ space such that the
 normal vector at the points in the intersections is included in the
 planes. The explicit form of the geodesic in AdS$_3$, expressed
 in the ambient $\vec{X}
 \in \mathbb{R}^{2,2}$ space,
 is \be
 \vec{X}=\f{R}{\s{\ap^2-1}}\sinh(\lambda/R)\cdot \vec{x}
 +R\left[\cosh(\lambda/R)-\f{\ap}{\s{\ap^2-1}}\sinh(\lambda/R)\right]
 \cdot \vec{y}, \label{geodesic}   \ee
 where $\ap=1+2\sinh^2\rho_0\sin^2(\pi l/L)$; $x$ and $y$ are
 defined by \ba \vec{x}&=&(\cosh\rho_0\cos t,\cosh\rho_0 \sin t
 ,\sinh\rho_0,0),\no  \vec{y}&=&\left(\cosh\rho_0\cos t,\cosh\rho_0 \sin t
 ,\sinh\rho_0\cos(2\pi l/L),\sinh\rho_0\sin(2\pi l/L)\right).\label{xy}\ea
 The length of the
 geodesic can be found as \be {\rm
Length}=\int ds=\int
 d\lambda=\lambda_*, \label{length}\ee where $\lambda_*$ is defined by
 \be \cosh(\lambda_*/R)=1+2\sinh^2\rho_0~\sin^2\f{\pi l}{L}.
 \label{dell} \ee
 Assuming that the UV cutoff energy is large $e^{\rho_0}\gg 1$, we can
 obtain the entropy (\ref{arealaw}) as follows
 (using (\ref{centralads}))
 \be S_A\simeq\f{R}{4G^{(3)}_N}\log\left(e^{2\rho_0}\sin^2\f{\pi l}{L}\right)
 =\f{c}{3}\log\left(e^{\rho_0}\sin\f{\pi l}{L}\right).\label{entropygamma}
 \ee Indeed, this
 entropy exactly coincides with the known 2D CFT result (\ref{entropyone}),
 including the (universal) coefficients
 after we remember the relation (\ref{delti}).

\subsection{Calculations in Poincare Coordinates}
\label{calc in Poincare}
\hspace{5mm} It is useful to repeat the similar analysis in the
Poincare coordinates (\ref{Poincare}). We pickup the spacial region
(again call $A$) $-l/2\leq x\leq l/2$ and consider its entanglement
entropy as in section \ref{deriv EE in inf sys}.
We can find the geodesic line
$\gamma_{A}$ between $x=-l/2$ and $x=l/2$ for a fixed time $t_0$
(see also later analysis in section \ref{AdS d+2/CFT d+1})
 \be (x,z)=\f{l}{2}(\cos s,\sin s),\ \ \
\ (\ep\leq s \leq \pi-\ep). \label{geodesicpp}\ee The infinitesimal
$\ep$ is the UV cutoff and leads to the cutoff $z_{UV}$ as
$z_{UV}=\f{l\ep}{2}$. Since $e^{\rho}\sim x^i/z$
near the boundary, we find $z\sim a$. The length of $\gamma_{A}$
can be found as \be {\rm Length}(\gamma_A)=2R\int^{\pi/2}_\ep \f{ds}{\sin
s}=-2R\log(\ep/2)=2R\log\f{l}{a}.\label{lenghthpo} \ee Finally the
entropy can be obtained as follows \be S_A=\f{{\rm
Length}(\gamma_A)}{4G^{(3)}_N}=\f{c}{3}\log\f{l}{a}.\label{entrppp}\ee This again
agrees with the well-known result (\ref{simple}) as expected.

\subsection{Entropy on Multiple Disjoint Intervals}
\label{EE on Multiple} \hspace{5mm} Next we proceed to more
complicated examples. Assume that the system $A$ consists of
multiple disjoint intervals. The entanglement entropy can be
computed as in (\ref{entamiult}). In the dual AdS$_3$ description,
the region $A$ corresponds to $\theta\in \cup_{i=1}^N[\f{2\pi
r_i}{L},\f{2\pi s_i}{L}]$ at the boundary. In this case it is not
straightforward to speculate the holographic screen (or minimal
surface) $\gamma_A$ . However, the result in the $1+1$ dimensional
conformal field theory (\ref{entamiult}) can be rewritten into the
following simple form \be S_A=\f{1}{4G_N^{(3)}}\left[\sum_{i,j}{\rm
Length}(r_j,s_i)- \sum_{i<j}{\rm Length}(r_j,r_i)-\sum_{i<j}{\rm
Length}(s_j,s_i)\right], \label{multads} \ee where ${\rm
Length}(A,B)$ denotes the length of the geodesic line between two
boundary points $A$ and $B$. This shows how we choose $\gamma_A$. It
is a linear combination of geodesic lines. Their coefficients are
either $1$ or $-1$. Thus some of the coefficients turn out to be
negative \footnote{ One may think the presence of minus signs is
confusing from the viewpoint of holographic screen. Instead we would
like to regard this as a singular (or just complicated) behavior
which is typical only in the lowest dimension. In higher dimensional
cases, we do not seem to have such a problem when $\partial A$ is
compact. Notice also that the total sum (\ref{multads}) is always
positive. If we replace the surface $\gamma_A$ with D-branes or
fundamental strings (remember the similarity to Wilson loops) , the
minus sign is analogous to ghost branes introduced recently in
\cite{ghost}.}. It is easy to see such negative coefficients are
necessary by considering the limit where $s_{i}$ coincides with
$r_{i+1}$ and requiring it reproduces the result for $N-1$
intervals.

\subsection{Finite Temperature Cases}
\hspace{5mm} Next we consider how to explain the entanglement
entropy (\ref{entropytemp}) at finite temperature $T=\beta^{-1}$
from the viewpoint of AdS/CFT correspondence. Since we assumed that
the spacial length of the total system $L$ is infinite, we have
$\beta/L\ll 1$. In such a high temperature circumstance, the gravity
dual of the conformal field theory is described by the Euclidean BTZ
black hole \cite{BTZ}. Its metric looks like \be
ds^2=(r^2-r_+^2)d\tau^2+ \f{R^2}{r^2-r^2_{+}}dr^2+r^2 d\vp^2.
\label{btzmet}\ee The Euclidean time is compactified as
$\tau\sim\tau+\f{2\pi R}{r_+}$ to obtain a smooth geometry. We also
impose the periodicity $\vp\sim \vp+2\pi$. By taking the boundary
limit $r\to \infty$, we find the relation between the boundary CFT
and the geometry (\ref{btzmet}) \be \f{\beta}{L}=\f{R}{r_{+}}\ll 1.
\label{relationbtz}\ee

The subsystem for which we consider the entanglement entropy is
given by $0\leq \vp\leq 2\pi l/L$ at the boundary. Then by extending
our proposal (\ref{arealaw}) to asymptotically AdS spaces,
 the entropy can be computed from the
length of the space-like geodesic starting from $\vp=0$ and ending
to $\vp=2\pi l/L$ at the boundary $r=\infty$ for a fixed time. To
find the geodesic line, it is useful to remember that the Euclidean
BTZ black hole at temperature $T$ is equivalent to thermal AdS$_3$
at temperature $1/T$. This equivalence can be interpreted as a
modular transformation in the boundary CFT \cite{MS}. If we define
the new coordinates \be r=r_{+}\cosh\rho,\ \ \
\tau=\f{R}{r_+}\theta,\ \ \ \vp=\f{R}{r_+}t,\label{newcor}\ee then
the metric (\ref{btzmet})\ indeed becomes the one in the Euclidean
Poincare coordinates with $t$ replaced by $it$. Now the computation
of the geodesic line is parallel with what we did in
section \ref{geodesics in AdS3}.
We only need to replace $\sinh\rho$ and $\sin t$ with
$\cosh\rho$ and $\sinh t$. In the end we find (\ref{length}) with
$\lambda_*$ is now given by \be \cosh\left(\f{\lambda_*}{R}\right)
=1+2\cosh^2\rho_0\sinh^2\left(\f{\pi
l}{\beta}\right),\label{delltwo} \ee where we took into account the
UV cutoff $e^{\rho_0}\sim \beta/a$. Then our area law
(\ref{arealaw}) precisely reproduces the known CFT result
(\ref{entropytemp}). We can extend these arguments to the multi
interval cases as in the zero temperature case. We again obtain the
formula (\ref{multads}) from the CFT result (\ref{entamiultem}).

\begin{figure}
\begin{center}
\includegraphics[height=3cm,clip]{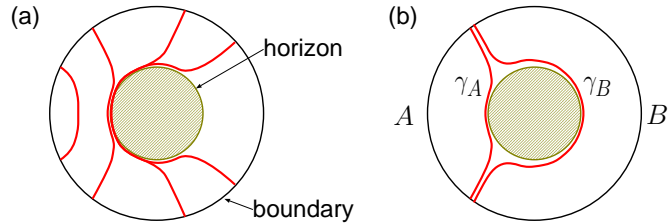}
\end{center}
\caption{ \label{fig: ads_blackhole.eps} (a) Minimal surfaces
$\gamma_A$ in the BTZ black hole for various sizes of $A$. (b)
$\gamma_A$ and $\gamma_B$ wrap the different parts of the horizon.}
\end{figure}

It is also useful to understand these calculations geometrically.
The geodesic line in the BTZ black hole takes the form shown in
Fig.\ \ref{fig: ads_blackhole.eps}(a). When the size of $A$ is
small, it is almost the same as the one in the ordinary AdS$_3$. As
the size becomes large, the turning point approaches the horizon and
eventually, the geodesic line covers a part of the horizon. This is
the reason why we find a thermal behavior of the entropy when
$l/\beta\gg 1$ in (\ref{thermalth}). The thermal entropy in a
conformal field theory is dual to the black hole entropy in its
gravity description via the AdS/CFT correspondence. In the presence
of a horizon, it is clear that $S_A$ is not equal to $S_B$ (remember
$B$ is the complement of $A$) since the corresponding geodesic lines
wrap different parts of the horizon (see Fig.\ \ref{fig:
ads_blackhole.eps}(b)). This is a typical property of entanglement
entropy at finite temperature as we mentioned in section
\ref{basics}.

\subsection{Massive Deformation}
\label{massive deform} \hspace{5mm} Now we would like to turn to
$1+1$ dimensional massive quantum field theories. Such a theory can
be typically obtained by perturbing a conformal field theory by a
relevant perturbation. In the dual gravity side, this corresponds to
a deformation of AdS$_3$ space. Since in the high energy limit the
mass gap can be ignored, the deformation only takes place for small
values $z<z_{IR}$ of $z$ in the Poincare coordinates. As in the
well-known examples \cite{KlSt,PoSt,MaNu} in AdS$_5$, we expect the
massive deformation caps off the end of the throat region.

Consider an $1+1$ dimensional infinite system divided into two
semi-infinite pieces and define the subsystem $A$ by one of them
(i.e. $A=A_{SL}$). Let us compute the entanglement entropy $S_A$ in
this setup. The important quantity in the massive theory is the
correlation length $\xi$. This is identified with $\xi\sim z_{IR}$
in the dual gravity side. Since we assumed that the subsystem $A$ is
infinite, we should take a geodesic (\ref{geodesicpp}) with a large
value of $l(\gg\xi)$. The geodesic starts from the UV cutoff $z=a$
and ends at the IR cutoff $z=\xi$. Then we obtain the length of this
geodesic as follows \be {\rm
Length}(\gamma_A)=\int^{2\xi/l}_{\ep=2a/l}\f{ds}{\sin
s}=R\log\f{\xi}{a}.\label{geodmass} \ee In the end we find its
entropy \be S_A=\f{{\rm
Length}(\gamma_A)}{4G^{(3)}_N}=\f{c}{6}\log\f{\xi}{a}.\label{entrpmas}
\ee This perfectly reproduces the known result (\ref{massivee}) in
the $1+1$ dimensional quantum field theory.

\section{Entanglement Entropy in CFT$_{d+1}$ from AdS$_{d+2}$}
\label{AdS d+2/CFT d+1}
\setcounter{equation}{0} \hspace{5mm} Since we have confirmed the
proposed relation (\ref{arealaw}) in the lowest dimensional case
$d=1$,  the next step is to examine higher dimensional cases. Our
proposal (\ref{arealaw})  argues that the entanglement entropy in
$d+1$ dimensional conformal field theories can be computed from the
area of the minimal surfaces in AdS$_{d+2}$ spaces. In the most of
arguments in this section
 we employ the Poincare coordinates (\ref{Poincare}) for simplicity.
Even though we cannot fully check our proposal due to the lack of
general analytical results in the CFT side, we will manage to obtain
some supporting evidences employing the previous results in section
\ref{(d+1)D CFT}.

\begin{figure}
\begin{center}
\includegraphics[width=8cm,clip]{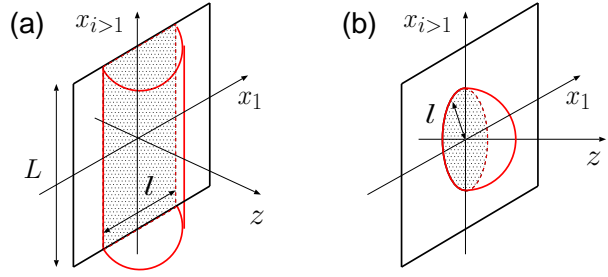}
\end{center}
\caption{
\label{fig: min_surf}
Minimal surfaces in
AdS$_{d+2}$: (a) $A_S$
and (b) $A_D$.
}
\end{figure}

\subsection{General Results}
\label{general results} \hspace{5mm} For specific choices of the
subsystem (or submanifold) $A$, it is easy to evaluate the area of
minimal surfaces directly in AdS$_{d+2}$ spaces of general
dimensions $d$. Essentially this is possible by applying the
techniques employed to compute the Wilson loops from AdS/CFT duality
\cite{Wilsonline,BeMa,GrWi}.

\subsubsection{Entanglement Entropy for Straight Belt $A_S$}
\label{EE for A_S}
\hspace{5mm} First consider the entanglement entropy for the
straight belt $A_S$ (\ref{straighta}) with the width $l$. The $d$
dimensional minimal surface in AdS$_{d+2}$ is given by minimizing
the area functional (we set $x=x_1$ in the coordinate system
(\ref{Poincare})) \be \mbox{Area}=R^dL^{d-1}\int^{l/2}_{-l/2}
dx\f{\s{1+(\f{dz}{dx})^2}}{z^d}.\ee Regarding $x$ as a time, we can
find the Hamiltonian which does not depend on $x$. This leads to \be
\f{dz}{dx}=\f{\s{z^{2d}_*-z^{2d}}}{z^d},\label{integgr} \ee where
$z_*$ is a constant. This equation determines the minimal surface
$\gamma_A$ (see Fig.\ \ref{fig: min_surf}(a)). Since $z=z_*$ is the
turning point of the minimal surface, we require\footnote{We
employed the formula $\int^1_0 dx x^{\mu-1}(1-x^\lambda)^{\nu-1}
=\f{B(\mu/\lambda,\nu)}{\lambda}$, where
$B(x,y)=\Gamma(x)\Gamma(y)/\Gamma(x+y)$.} \be
\f{l}{2}=\int^{z_*}_{0}dz \f{z^d}{\s{z^{2d}_*-z^{2d}}}
=\f{\s{\pi}~\Gamma(\f{d+1}{2d})}{\Gamma(\f{1}{2d})}z_*.\label{reqgg}
\ee

Then the area is given by \be {\rm
Area}_{A_S}=\f{2R^d}{d-1}\left(\f{L}{a}\right)^{d-1}-2IR^d
\left(\f{L}{z_*}\right)^{d-1},\label{aregne} \ee where $I$ is the
constant \be
I=\f{1}{d-1}-\int^{1}_{0}\f{dy}{y^d}\left(\f{1}{\s{1-y^{2d}}}-1\right)
=-\f{\s{\pi}~\Gamma(\f{1-d}{2d})}{2d~\Gamma(\f{1}{2d})}>0.
\label{consti}\ee

In the end, we find the entanglement entropy from (\ref{arealaw})
using (\ref{reqgg}), (\ref{aregne}) and (\ref{consti})

\be
S_{A_{S}}=\f{1}{4G^{(d+2)}_{N}}\left[\f{2R^d}{d-1}\left(\f{L}{a}\right)^{d-1}
-\f{2^d \pi^{d/2} R^{d}}{d-1}\left(\f{\Gamma(\f{d+1}{2d})}
{\Gamma(\f{1}{2d})}\right)^{d} \left(\f{L}{l}\right)^{ d-1}\right],
\label{areaone} \ee

Notice that the first divergent term is proportional to the area of
$\de A$ i.e. $L^{d-1}$ as we expect from the known area law in the
field theory computations (\ref{divarea}). The second term is finite
and thus is universal (i.e. does not depend on the cutoff). This is
the quantity which we can directly compare with the field theory
counterpart. Notice that our result (\ref{areaone}) does not include
subleading divergent terms ${\mathcal{O}}(a^{-d+3})$.  This is
because $A_S$ is in the straight shape. When we deform and bend it,
the subleading divergent terms appear in general as we will see
later in another example. For example, in the 4D case, the absence
of $\log$ term is clear from the previous CFT analysis (\ref{aceq}).

\subsubsection{Entanglement Entropy for Circular Disk $A_D$}
\label{EE for A_D}
\hspace{5mm} Next we examine the case where subsystem $A$ is given
by the circular disk $A_D$ (radius $l$) as defined in (\ref{diska}).
We use the polar coordinate for ${\mathbb{R}}^{d}$ such that
$\sum_{i=1}^d dx_i^2=dr^2+r^2d\Omega^2_{d-1}$. The minimum surface
is the $d$ dimensional ball B$^d$ defined by $z=z(r)$ (and
$\Omega_{d-1}$ takes arbitrary values). The function $z(r)$ is found
by minimizing the area functional \be \mbox{Area}_{A_D}=R^d \cdot
\mbox{Vol}({\mbox{S}}^{d-1}) \cdot\int_0^l dr
r^{d-1}\f{\s{1+(\f{dz}{dr})^2}}{z^d} . \label{areafuncd}\ee We can
find the following simple solution from the equation of
motion\footnote{The equation of motion is given by
$rzz''+(d-1)z(z')^3+(d-1)zz'+dr(z')^2+dr=0.$}
 for (\ref{areafuncd})
 \be r^2+z^2=l^2.\label{solgen}\ee
 Thus $\gamma_A$ is a half of a $d$ dimensional sphere (see
 Fig.\ \ref{fig: min_surf}(b)).
This can be also found from the conformal map of the simplest case
where $\de A$ is a single straight line (i.e. $A=A_{SL}$) into
$A_D$. Then we obtain its area
\begin{eqnarray}
{\rm Area}_{A_D}
&=&
\mbox{Vol}({\mbox{S}}^{d-1}) \cdot R^d\cdot
\int^1_{a/l}dy \f{(1-y^2)^{(d-2)/2}}{y^d}
\nonumber \\
&=& \f{2\pi^{d/2}R^d}{\Gamma(d/2)}\cdot
\left[\f{1}{d-1}\left(\f{l}{a}\right)^{d-1}-\f{d-2}{2(d-3)}
\left(\f{l}{a}\right)^{d-3} +\ddd\right].\label{aregencir}
\end{eqnarray}
In
this expression (\ref{aregencir}), the omitted subleading terms
$\ddd$ of the order ${\mathcal{O}}(a^{-d+5})$ include the
logarithmic term $\sim \log\f{l}{a}$ when $d$ is odd. On the other
hand, if $d$ is even, the series end up with a constant term. Taking
into account these, the final expression of the entanglement entropy
can be found as follows applying (\ref{arealaw})
\begin{eqnarray}
S_{A_{D}} &=&
\f{2\pi^{d/2}R^d}{4G^{(d+2)}_{N}\Gamma(d/2)}
\int^1_{a/l}dy \f{(1-y^2)^{(d-2)/2}}{y^d}  \nonumber \\
&=&
  p_1  \left(l/a\right)^{d-1}
+ p_3 \left(l/a \right)^{d-3}
+\cdots  \label{areatwo}  \\
&&
\cdots +\left\{
\begin{array}{ll}
\displaystyle p_{d-1}\left(l/a\right) + p_d + \mathcal{O}(a/l), &
 \mbox{$d$: even},   \\
\displaystyle p_{d-2} \left(l/a\right)^{2} + q \log
\left(l/a\right)+ \mathcal{O}(1),
&  \mbox{$d$: odd,}   \\
\end{array}
\right.
 \nonumber
\end{eqnarray}
where the coefficients are defined by \ba p_1/C &=& (d-1)^{-1},\ \
p_3/C = - (d-2)/[2(d-3)],\ \ \ddd \no p_d/C &=&
(2\s{\pi})^{-1}\Gamma(d/2) \Gamma\left((1-d)/2\right) \ \
(\mbox{if}\ \  d=\mbox{even}),\no q/C &=&
(-)^{(d-1)/2}(d-2)!!/(d-1)!! \ \  (\mbox{if}\ \  d=\mbox{odd}),
\no
&&\mbox{where}\ \ \  C\equiv
\f{\pi^{d/2}R^d}{2G^{d+2}_{N}\Gamma(d/2)}.  \label{entropydg}\ea

We notice that the result (\ref{entropydg}) includes a leading UV
divergent term $\sim a^{-d+1}$ and its coefficient
 is proportional to the area of the boundary $\de A$ as expected from
the area law \cite{Bombelli,Srednicki} in the field theories
(\ref{divarea}). We have also subleading divergence terms which
reflects the form of the boundary $\de A$.

In particular, we prefer a physical quantity that is independent of
the cutoff (i.e.~universal). The final term in (\ref{entropydg}) has
such a property. When $d$ is even, it is given by a constant $p_d$.
This seems to be somewhat analogous to the topological entanglement
entropy (or quantum dimension) recently introduced in $2+1$ D
topological field theories \cite{Kitaev05, Levin05}, though our
theory is not topological.

On the other hand, when $d$ is odd, the coefficient of the
logarithmic term $\sim \log(l/a)$ is universal as was so in the 2D
case (\ref{simple}). Indeed, we found such a term in the analysis of
4D conformal field theories e.g. (\ref{fouren}), which is
proportional to the central charge. This issue will also be
discussed in detail later.

This result is based on an explicit calculation when $A=A_D$.
However, from the paper \cite{GrWi}, we find that the behavior
(\ref{areatwo}) is also true for any compact submanifold $A$ with
different coefficient $p_k$ and $q$ depending on the shape of $A$.

\subsubsection{Multiple Loops}
\label{multiple loops} \hspace{5mm} When the system $A$ consists of
$M$ disconnected regions (we call them $A_1,A_2,\ddd,A_M$, we need
to find the minimal surface $\gamma_A$ whose boundary $\de A$ is
$A_1\cup A_2\cup\ddd\cup A_M$. If the distance between $A_i$s are
small enough we may find a connected surface of this property.
However, if they are far apart, $\gamma_A$ can be separated into
several pieces as pointed out by \cite{GrOo} in the analogous
problem of Wilson loop computations. Even if we take into account
this complexity, the inequality (subadditivity) $S(A)\leq
S(A_1)+S(A_2) +\ddd +S(A_N)$ is clearly satisfied.

It is also useful to consider a singular limit of such a multiple
component case, i.e. when the subsystem $A$ consists of the multiple
straight belts $A_{S(1)},A_{S(2)},\ddd , A_{S(N)}$. In this
situation, we can naturally obtain the entanglement entropy from the
formula (\ref{multads}) by replacing the geodesic distance with the
area of the minimal surfaces. This agrees with the free field
computation which is a straightforward generalization of the result
in section \ref{rougth estimate}.

\subsection{Entanglement Entropy in ${\mathcal N}=4$ SYM from AdS$_5\times$ S$^5$}
\label{EE in SYM from AdS/CFT}
\hspace{5mm} So far we have discussed low energy gravity theories on
 AdS$_{d+2}$ and have not been careful about its high energy
completion as quantum gravity. To understand the holographic
relation better including the various quantum corrections, it is
necessary to realize a concrete embedding into string theory. The
most important such example is the AdS$_5\times $S$^5$ background in
type IIB string theory. This background preserves the maximal 32
supersymmetries and is considered to be dual to the ${\mathcal N}=4$
$SU(N)$ Super Yang-Mills theory \cite{Maldacena}. The supergravity
approximation corresponds to the large t' Hooft coupling
$\lambda=Ng_{YM}^2\gg 1$ (i.e. strongly coupled) region. The planar
limit $N\to\infty$ is equivalent to the weakly coupled region
$g_s\to 0$ of type IIB string. Since we perform the supergravity
analysis, the dual gauge theory is strongly coupled and the large
$N$ limit is taken.

The 5D Newton constant $G^{(5)}_N$ is given in terms of the 10D one
\be G^{(10)}_N=\f{\kappa^2}{8\pi}=8\pi^6\al^4
g_s^2,\label{gnewten}\ee as follows \be
G^{(5)}_N=\f{G^{(10)}_N}{R^5{\rm Vol}({\mbox{S}}^5)}=
\f{G^{(10)}_N}{\pi^3 R^5}.\label{gnewfive}\ee The radius $R$ of
AdS$_5$ and $S^5$ is expressed as \be R=(4\pi g_s \al^2
N)^{\f{1}{4}}.\label{rdef} \ee

Plugging these values (\ref{gnewfive}) and (\ref{rdef}) into the
previous results (\ref{areaone}) and (\ref{areatwo}), we obtain the
following prediction of entanglement entropies in ${\mathcal N}=4$
$SU(N)$ super Yang-Mills theory

\ba S_{A_S}&=&\f{N^2L^2}{2\pi a^2}-2\s{\pi}
\left(\f{\Gamma\left(\f23\right)}{\Gamma\left(\f16\right)}
\right)^3\f{N^2L^2}{l^2}, \label{entrodthe} \\
S_{A_D}&=&N^2\left[\f{l^2}{a^2}-\log\left(\f{l}{a}\right)
+{\mathcal{O}(1)}\right]\label{entradfivecir}.\ea Notice that these
are proportional to $N^2$ as expected since the number of fields in
the $SU(N)$ gauge theory is proportional to $N^2$. Interestingly,
(\ref{entrodthe}) does not depend on $g_{YM}^2=\f{g_s}{2\pi}$.

Let us examine the first result (\ref{entrodthe}). We notice that it
has the same functional form as in the free field theories
(\ref{exactst}). Since the second term in (\ref{entrodthe}) is
finite, it is interesting to compare its coefficient with that of
the free field theory result. The finite term in (\ref{entrodthe})
is numerically expressed as \be
S^{Sugra}_{A_S}|_{finite}\simeq-0.0510\cdot
\f{N^2L^2}{l^2}.\label{numeentr}\ee On the other hand, in the free
field theory side we can employ the estimations (\ref{eq: k for
boson and fermion for generic D no approx}). The ${\mathcal N}=4$
super Yang-Mills consists of a gauge field $A_{\mu}$, six real
scalar fields $(\phi^1,\phi^2,\ddd,\phi^6)$ and four Majorana
fermions $(\psi^1_\ap,\psi^2_\ap,\psi^3_\ap,\psi^4_\ap)$. As we
explained in section \ref{EE of 4D gauge field}, the contribution
from the gauge field is the same as those from two real scalar
fields. In this way the total entropy in the free Yang-Mills theory
is the same as those from 8 real scalars and 4 Majorana fermions .
Thus we obtain from (\ref{eq: k for boson and fermion for generic D
no approx}) the following estimation \be S^{Free
YM}_{A_S}|_{finite}\simeq -(8\times 0.0049+4\times 0.0097)\cdot
\f{N^2L^2}{l^2}=-0.078 \cdot \f{N^2L^2}{l^2}.\label{numeentrr}\ee We
observe that the free field result is larger than that in the
gravity dual by a factor $\sim\f{3}{2}$. This deviation is expected
since the computation of the entanglement entropy\footnote{ As we
notice in section \ref{EE and central charge, 4D},
some parts of entanglement entropy are
proportional to the central charges. They remain the same under
exactly marginal deformation (e.g. changing coupling $g_{YM}$) since
central charges do so.} includes non-BPS quantities due to the
anti-periodic boundary condition of fermions which appears when we
compute the partition function on $n-$sheeted manifold $M_n$. This
situation is very similar to the computation of thermal entropy
\cite{GKP}, where we have a similar discrepancy (so-called
$\f{4}{3}$ problem). The fact that the discrepancy is of order one
also in our computation can be thought as an encouraging evidence
for our proposal. Also notice that the coefficient in the free
Yang-Mills is larger than the one in the strongly coupled
Yang-Mills. This is natural since the interaction of the form
Tr$[\phi_i,\phi_j]^2$ reduces the degrees of freedom
\cite{adsreview}.

Next we turn to our second result (\ref{entradfivecir}). In addition
to the area law divergence, it includes a logarithmic term, whose
coefficient is universal. This qualitative dependence of the entropy
(\ref{entradfivecir}) on $l$ agrees with our previous result from
the Weyl anomaly (\ref{fouren}). We will discuss the coefficient in
front of the logarithmic term in more detail in the next subsection.

\subsection{Entanglement Entropy  and Central Charges in 4D CFT
from AdS$_5$}
\label{EE 4D CFT}

\hspace{5mm} We can extend the previous computation to more general
(i.e. less supersymmetric) conformal backgrounds by replacing S$^5$
with a compact five dimensional Einstein manifold $X_5$. The radius
$R$ of AdS$_5$ and $X_5$ is given by \cite{HeSk,Gubser} \be
R=\left(\f{4\pi^4g_s\al^2 N}{\mbox{Vol}(X_5)}\right)^{\f14},\ee
where $N$ is again the number of D3-branes (or rank of the gauge
group). The volume $\mbox{Vol}(X_5)$ of $X_5$  is known to be
inversely proportional to the central charge $a$ \cite{Gubser}. Note
that $a=c$ always holds when a CFT has its gravity dual of the form
AdS$_5\times X_5$. In the ${\mathcal N=4}$ $SU(N)$ super Yang-Mills
theory the central charge is given by
$a_{{\mathcal{N}}=4}=\f{N^2-1}{4}\simeq \f{N^2}{4}$.

The entanglement entropy $S_A$ in general 4D CFTs of this type is
given in terms of $S^{{\mathcal{N}}=4}_A$ in ${\mathcal N=4}$
$SU(N)$ super Yang-Mills theory \be
S_A=\left(\f{a}{a_{{\mathcal{N}}=4}}\right)\cdot
S^{{\mathcal{N}}=4}_A, \label{acent} \ee i.e. $S_A$ is proportional
to the central charge $a$. This is naturally understood by
considering that the central charge $a$ measures degrees of freedom
in the 4D CFT. Notice that here we are assuming a strongly coupled
4D CFT in order to apply the AdS/CFT duality. In our previous CFT
analysis done in section \ref{EE and central charge, 4D},
we have only shown that a part of
entanglement entropy is proportional to the central charge
(\ref{fouren}).

As we have seen, the coefficient of the logarithmic term (called
$\gamma_2$ in section \ref{EE and central charge, 4D}) in
(\ref{entradfivecir}) is universal and is given by the central
charge $a$ times a numerical factor. Thus it is very interesting to
compare the factor between the gauge theory and the gravity. When
the 2D surface $\de A$ is generic (with a finite size) and the
background is an arbitrary asymptotically AdS$_5$ space, the
logarithmic term in the area of the minimal surface $\gamma_A$ can
be found from the general formula given in \cite{GrWi}. This leads
to \ba l\f{d\mbox{Area}(\de A)}{dl}|_{finite}=\int_{\de A}d^2
x\s{g}\left(-\f{1}{8}|H|^2
-\f{1}{4}g^{\ap\beta}R_{\ap\beta}+\f{1}{12}R\right),
\label{adsel}\ea where $\ap$ and $\beta$ are the coordinates which
are tangent to $\de A$ and orthogonal to $i,j$ directions; $H$ is
the mean curvature. As we did in section \ref{EE and central charge,
4D}, we work below under the special assumption that the second
fundamental forms are zero to make arguments simple. Then we can
show \ba R&\simeq &R_{\Sigma=\de A}+2R_{ii}-R_{ijij}, \no
g^{\ap\beta}R_{\ap\beta}&\simeq &R_{\Sigma=\de
A}+R_{ii}-R_{ijij}.\ea We can also neglect $|H|^2$ term in
(\ref{adsel}). In the end, we can rewrite (\ref{adsel}) into the
following form \be l\f{d\mbox{Area}(\de A)}{dl}|_{finite}=\int_{\de
A}d^2 x\s{g}\left(\f{1}{6}R_{ijij}-\f{1}{12}R_{ii}
-\f{1}{6}R_{\Sigma=\de A}\right). \label{gravrac} \ee By considering
the setup dual to the 4D $\mathcal{N}=4$ $SU(N)$ super Yang-Mills
theory ($a=c\simeq\f{N^2}{4}$), it is straightforward to check that
the gravity result (\ref{gravrac}) agrees\footnote{Under this
assumption we cannot deal with the circular disk case $A_D$ because
the second fundamental forms are non-zero (i.e.
$\Gamma^{i}_{\ap\beta}\neq 0$). However, it is possible to see that
the contribution (\ref{topt}), which comes from the topological term
$\ti{R}^2$, coincides with the gravity result. Indeed we expect that
the other contribution from the Weyl tensor term $W^2$ is vanishing
since the disk is conformally equivalent to the straight line, in
which case there is no log term (notice also that the $W^2$ term is
a conformal invariant).} with the previous result (\ref{aceq})
obtained from the Weyl anomaly\footnote{ The derivation of the Weyl
anomaly from the AdS/CFT duality was first done in \cite{HeSk}.}. It
will be an interesting future problem to examine terms which include
the second fundamental forms and check the complete agreement.

\subsection{Entanglement Entropy from AdS$_{4,7}\times $S$^{7,4}$ in M-theory }
\label{EE from AdS_47}
\hspace{5mm} Other important supersymmetric examples of AdS spaces
are AdS$_{4}\times $S$^{7}$ and AdS$_{7}\times $S$^{4}$ in
eleven dimensional supergravity (or M-theory). They preserve the
maximal 32 supersymmetries. They are considered to be dual to 3D
${\mathcal N}=8$ SCFT and 6D $(2,0)$ SCFT, respectively
\cite{Maldacena,adsreview}.
 They are obtained from the world-volume theories on M2 and M5-branes
(or strongly coupled limit of D2 and D4-branes). The numbers of the branes
are denoted by $N$. Since these theories have not been
completely understood due to the strongly coupled problem,
it will be very useful to compute any new physical quantities.

The 11D Newton constant  $G^{(11)}_{N}$ is given in terms of  11D
plank length $l_p$ as follows\footnote{Our convention is such that
$S_{11D sugra}=\f{1}{2\kappa_{11}^2}\int d^{11}x[\s{g}R+\cdots]$. We
follow the convention in \cite{adsreview}.} \be (2\pi)^8l_p^9=16\pi
G^{(11)}_{N}=2\kappa_{11}^2. \label{kappepl} \ee

Let us first discuss  the AdS$_{4}\times $S$^{7}$ example. The
radius of AdS$_4$ and $S^7$ are
\be
2R_{\mathrm{AdS}_4}=R_{S^7}=l_p(32\pi^2
N)^{\f16}.\label{radiuadsm}
\ee
The four dimensional Newton constant
can be found after the compactification on
 $S^7$ \be
G^{(4)}_N=\f{48\pi^3l_p^9}{R_{S^7}^7}.\label{newtonff} \ee Then we
find the following entanglement entropy  defined for the straight
belt  $A_S$ \be S_{A_S}=\f{{\rm
Area}}{4G^{(4)}_{N}}=\f{\s{2}}{3}N^{3/2}\left[\f{L}{a}-
\f{4\pi^3}{\Gamma(1/4)^4}\f{L}{l}\right].\label{entompne} \ee The
entropy for the circular disk $A_D$ we find \be S_{A_D}=\f{{\rm
Area}}{4G^{(4)}_{N}}=\f{\s{2}\pi}{3}N^{3/2}
\left[\f{l}{a}-1\right].\label{entompnee}\ee Notice that the
constant terms in (\ref{entompne}) and (\ref{entompnee}) are
universal. The dependence $\sim N^{3/2}$ of degrees of freedom is
typical in the 3D ${\mathcal N}=8$  SCFT.

In the AdS$_7\times$ S$^4$ case, in addition to (\ref{kappepl}), we
have
\be
R_{\mathrm{AdS}_7}=2R_{S^4}=2l_p(\pi
N)^{\f13},\label{radiuadsmse}
\ee and \be
G^{(7)}_N=\f{6\pi^5l_p^9}{R_{S^4}^4}. \label{newtonf}\ee Then we
find the following results  \ba
S_{A_S}&=&\f{2}{3\pi^2}N^{3}\left[\f{L^4}{a^4}-
16\pi^{5/2}\left(\f{\Gamma(3/5)}{\Gamma(1/10)}\right)^5\f{L^4}{l^4}\right].
\label{entompnemf} \\
S_{A_D}
&=&\f{32}{9}N^{3}
\left[\f{1}{4}\cdot\f{l^4}{a^4}-\f{3}{4}\cdot\f{l^2}{a^2}
+\f{3}{8}\log(l/a)\right].\label{entompneemf} \ea
Notice that the
constant term in (\ref{entompnemf}) and the coefficient of $\log(l/a)$ in
 (\ref{entompneemf}) are universal. The overall
 dependence $\sim N^3$ is again peculiar
 to 6D $(2,0)$ SCFT.

\subsection{Finite Temperature Case}
\label{finite temperature case} \hspace{5mm} Consider the
${\mathcal{N}}=4$ super Yang-Mills theory on $\mathbb{R}^4$ at
finite temperature $T$. This system is dual to the AdS black hole
geometry \cite{ADSWitten,Witten} \be
ds^2=R^2\left[\f{du^2}{hu^2}+u^2\left(-h
dt^2+dx_1^2+dx_2^2+dx_3^2\right)+d\Omega_5^2\right],\label{adsbl}\ee
where \be h=1-\f{u_0^4}{u^4},\ \ \ u_0=\pi T.\label{nonext} \ee
Various analyzes show that this theory has properties of a confining
gauge theory \cite{Witten,adsreview}.

We would like to compute the entanglement entropy for the straight
line in this model. The subsystem $A$ is defined by $t=\mathrm{fixed}$,
$-l/2\leq x(\equiv x_1)\leq l/2$, $u\to \infty$, and
$x_2,x_3=$arbitrary. The regularized volume in the $x_2$ and $x_3$
direction is denoted by $L^2$. Then the area is given by\footnote{
This system is very similar to the one that appears in the
computation of Wilson loop at finite temperature \cite{Wilsontemp}.}
\be {\rm Area}=R^3L^2\int^{l/2}_{-l/2}dx
u^3\s{1+\f{u'^2}{u^4-u_0^4}}.\label{arefint}\ee We can integrate the
equation of motion as \be
\f{du}{dx}=\s{(u^4-u_0^4)(u^6/u_*^6-1)}.\label{intgege}\ee We
require \be \f{l}{2}=\int^{\infty}_{u_*}du
\f{1}{\s{(u^4-u_0^4)(u^6/u_*^6-1)}},\label{lengthcond}\ee where
$u_*(>u_0)$ is the value of $u(x)$ at the turning point $x=0$. Using
(\ref{intgege}) we can rewrite (\ref{arefint}) \be {\rm
Area}=2R^3L^2\int^{\infty}_{u_*}du
\f{u^6}{\s{(u^4-u_0^4)(u^6-u_*^6)}}. \label{arefintt}\ee

As usual, (\ref{arefintt}) contains the UV divergent term which is
proportional to $a^{-2}$. However, we are interested in the term
which is peculiar to this kind of confining gauge theory. Indeed we
can find that in the large $l$ limit (i.e. $u_*\sim u_0$), the main
contribution (except the UV divergence) of the integrals
(\ref{lengthcond}) and (\ref{arefintt}) comes from the region near
$u=u_*$, which leads to the relation \be {\rm Area}_{finite}\sim
\pi^3 N^2 R^3L^2lT^3.\label{relationgg} \ee Thus we obtain the
finite part (i.e. we subtracted the UV divergent terms) of the
entropy in this limit \be
S_{finite}=\f{\pi^2N^2}{2}T^3L^2l=\f{\pi^2N^2}{2}T^3\times {\rm
Area}(A).\label{entrofindd} \ee The important point is that this
entropy (\ref{entrofindd}) is proportional to the area of not $\de
A$ but $A$ as opposed to the area law term (\ref{arealaw}). Thus it
is extensive as in the thermal entropy. This agrees with the field
theory side since the entanglement entropy should include the
thermal entropy contribution as is obvious from its definition. In
the gravity side, it occurs because $\gamma_A$ wraps a part of the
black hole horizon and thus (\ref{entrofindd}) is equal to the
fraction of black hole entropy, which shows the thermal behavior.
This means that the behavior of the entanglement entropy is rather
different before and after the cofiniment/de-confinement transition
when we consider the ${\cal{N}}=4$ super Yang-Mills on
${\mathbb{R}}\times \mbox{S}^3$ (see \cite{AMMPR} and references
therein for recent studies of this phase transition). Thus the
entanglement entropy plays an role similar to an ``order parameter''.
Other geometrical properties are also parallel with the AdS$_3$ case
as Fig.\ \ref{fig: ads_blackhole.eps} shows.

\subsection{Massive Deformations}
\label{massive deform 2} \hspace{5mm} As a final example we would
like to discuss the entanglement entropy in $d+1$ dimensional
massive QFTs.  Typically we can obtain such theories by considering
massive deformations of a $d+1$ dimensional CFT. In principle, this
can be done by looking at supergravity solutions dual to
non-conformal field theories such as \cite{PoSt,KlSt,MaNu}. Instead
here we approximate the geometry simply by cutting off the IR region
$z>\xi$ of the AdS$_{d+2}$ space as we did in the $d=1$ case. Here
$\xi$ is the correlation length and we are assuming $\xi\gg l$.
\footnote{ When a quantum ground state of a massive theory has
non-trivial Berry phases, contribution from the Berry phase to the
entanglement entropy is also important \cite{Ryu06}. }

\subsubsection{Straight Belt $A_S$}
\label{A_S}
\hspace{5mm} Let us start with the computation of the entanglement
entropy for the straight belt in a massive theory by the simple
method explained in the above. This leads to the following
estimation
\begin{eqnarray}
S_{A_S} & = & 2\f{R^d L^{d-1}}{4G^{(d+2)}_N}\times \int_{a}^{\xi} dz
\frac{ \sqrt{ \left( \frac{dx}{dz} \right)^2 + 1 }
}{z^d}
=
\frac{L^{d-1}R^d}{2G^{(d+2)}_N z_*^{d-1}} \int_{a/z_*}^{\xi/z_*}
\frac{d\lambda}{ \lambda^d \sqrt{1-\lambda^{2d}} } \no
&=& \frac{R^d L^{d-1}}{2G^{(d+2)}_N}
\Biggl[ - \frac{a^{-d+1}}{-d+1} +
\frac{\xi^{-d+1}}{-d+1}
\nonumber \\
&&\!\!\!\!\! \quad +\frac{1}{2}\frac{1}{d+1} \frac{\xi^{d+1}}{z_*^{2d}} +
\cdots + \frac{(2n-1)!!}{(2n)!!} \frac{1}{(2n-1)d+1}
\frac{\xi^{(2n-1)d+1}}{z_*^{2nd}} + \cdots \Biggr]
\nonumber \\
&=&\!\!\! \frac{R^d L^{d-1}}{2G^{(d+2)}_N} \left[
\f{a^{-d+1}}{d-1}\!-\f{ \xi^{-d+1}}{d-1} \!+ r_1
\frac{\xi^{d+1}}{{l^{2d}}}\! +\!\cdots\! +r_n
\frac{\xi^{(2n-1)d+1}}{{l^{2nd}}} \!+ \! \cdots \right],
\label{massivest}
\end{eqnarray}
where $r_i$s are some numerical constants. We assumed the same form
of the minimal surface as in the conformal case and thus the
relation between $z_*$ and $l$ is the same as before (\ref{reqgg}).

\subsubsection{Circular Disk $A_D$}
\label{A_D}
\hspace{5mm} Next we examine the entanglement entropy for the
circular disk $A_D$ (radius $l$) in a massive theory. We assume the
same minimal surface $r^2+z^2=l^2$ as in the conformal case.
\begin{eqnarray}
S_{A_D} &=& \f{R^{d}\mathrm{Vol}(S^{d-1})}{4G^{(d+2)}_N} \int dr\, r^{d-1}
\frac{\sqrt{1 + (\f{dz}{dr})^2}}{z^d} \no
&=&\f{2\pi^{d/2} R^d }{4\Gamma(d/2)G^{(d+2)}_N} \int_{a/l} ^{\xi/l}  dy \frac{
\left(1 - y^2 \right)^{(d-2)/2} }{y^d}. \label{massivedis}
\end{eqnarray}
The integral in the final expression in
  (\ref{massivedis}) has the following series expansion
when $d$ is even (we set $d=2n)$
\begin{eqnarray}
&& \int^{\xi/l}_{a/l}  dy\, \frac{(1-y^2)^{(d-2)/2}}{y^d}
\nonumber \\
&=& \Biggl[-\frac{1}{d-1} \left( \frac{l}{\xi} \right)^{d-1} +
\frac{d-2}{2(d-3)} \left( \frac{l}{\xi} \right)^{d-3}+\ddd
\nonumber \\
&& \cdots - \frac{(-1)^n}{2^n n!} \frac{ (d-2n)(d-2n+2)\cdots(d-2)
}{d-2n-1} \left( \frac{l}{\xi} \right)^{d-2n-1} +\cdots \Biggr]
\nonumber \\
&& +\Biggl[ \frac{1}{d-1} \left( \frac{l}{a} \right)^{d-1} - \frac{1}{2}
\frac{d-2}{d-3} \left( \frac{l}{a} \right)^{d-3} + \cdots \no
&&\ddd+
\frac{(-1)^n}{2^n n!} \frac{ (d-2n)(d-2n+2)\cdots(d-2) }{d-2n-1}
\left( \frac{l}{a} \right)^{d-2n-1} \Biggr] ,   \label{integcir}
\end{eqnarray}
where the expansion of $a/l$ is truncated since we
take the limit  $a\to 0$ in the final expression.

When $d$ is odd ($d=2n+1$), we obtain the same result
(\ref{integcir}) except that we have to be careful about the two
terms $\mathcal{O}\left((l/\xi)^{d-2n-1}\right)$ and
$\mathcal{O}\left((l/a)^{d-2n-1}\right)$ in  (\ref{integcir})  which
are proportional to $\f{1}{d-2n-1}\to \infty$. The divergences are
canceled out and produce a log term
\begin{eqnarray}
(-1)^{\f{d-1}{2}} \f{(d-2)!!}{(d-1)!!}~\log\f{\xi}{a}.  \label{loglogm}
\end{eqnarray}
Thus in the odd $d$ case, we just have to replace the two terms in
(\ref{integcir}) with (\ref{loglogm}). Note that this term has the
same coefficient as the one in the conformal case, i.e.  $q/C$ in
(\ref{entropydg}) in our approximation. In summary we find
\begin{eqnarray}
S_{A_{D}}  &=&\f{2\pi^{d/2} R^d }{4\Gamma(d/2)G^{(d+2)}_N}
 \Biggl[  \f{1}{d-1}\f{l^{d-1}}{a^{d-1}}+(\mbox{subleading divergences} \
{\mathcal{O}}(l^{d-3}/a^{d-3}))\Biggr] \no
&&+\f{2\pi^{d/2} R^d }{4\Gamma(d/2)G^{(d+2)}_N}
 \Biggl[ -\f{1}{d-1}\f{l^{d-1}}{\xi^{d-1}}+{\mathcal{O}} (l^{d-3}/\xi^{d-3})\Biggr]\no
&&
+\left\{
\begin{array}{ll}
\displaystyle
0 &
\ \ \  \mbox{$d$: even},   \\
\displaystyle
(-1)^{\f{d-1}{2}}\f{2\pi^{d/2} (d-2)!! R^d }{4\Gamma(d/2)(d-1)!! G^{(d+2)}_N} \cdot
\log\f{\xi}{a}
& \ \ \   \mbox{$d$: odd.}   \\
\end{array}
\right.   \label{finmassive}
\end{eqnarray}

\subsection{Entanglement Entropy in Some Non-Conformal Theories}

\hspace{5mm} The best way to derive the entanglement entropy in
massive (or non-conformal) theories is to start with their dual
supergravity backgrounds instead of the previous crude
approximation. Since usually such backgrounds include complicated
metric and many other fields, we would like to make a first step by
looking at some simple cases such as the near horizon limit of
Dp-branes. Here we would like to examine the example of the
D2-branes and NS5-branes. It will be an interesting future problem
to analyze more complicated but more realistic examples.

\subsubsection{D2-branes Case}

\hspace{5mm} The decoupling limit of supergravity solution for $N$
D2-branes is given by the following metric and dilaton \cite{IMSY}
 \ba ds^2&=&\al
\left(\f{U^{5/2}}{g_{YM}\s{6\pi^2N}}(-dx^2_0+dx^2_1+dx^2_2)+\f{g_{YM}\s{6\pi
N^2}}
{U^{5/2}}dU^2+\f{g_{YM}\s{6\pi^2N}}{U^{1/2}}(d\Omega_6)^2\right),\no
e^{2\phi}&=&g_{YM}^2\left(\f{g_{YM}\s{6\pi^2N}}{U^{5/2}}\right)^{1/2}.
\label{sugradt} \ea

This supergravity background (\ref{sugradt}) is dual to the field
theory limit of world-volume theory on $N$ D2-branes. This field
theory is described by the 3D $SU(N)$ super Yang-Mills theory with
the dimensionful coupling constant $g_{YM}(\propto
\mbox{energy}^{1/2})$. The radial direction $U$ is proportional to
the energy scale in this field theory.

To avoid the strongly coupled region $g_s\gg 1$ and the high
curvature region $\al R\gg 1$, we trust the supergravity solution
(\ref{sugradt}) when \cite{IMSY} \be g_{YM}^2N^{1/5}\ll U\ll
g_{YM}^2N. \label{constdt}\ee The other two regions $U \gg
g_{YM}^2N^{1/5}$ and $U\ll g_{YM}^2N$ are well described by the 3D
superconformal field theory (or M2-branes) and the weakly coupled
Yang-Mills theory, respectively.

Under this condition (\ref{constdt}), we would like to compute the
entanglement entropy holographically in the straight belt case
$A=A_{S}$. First we notice that the dilaton is not constant and thus
the definition of $G_N^{(4)}$ is not clear. However, it is easy to
find a natural extension of our formula (\ref{arealaw}) by
remembering the relation
$\f{1}{G^{(4)}_N}=\f{1}{G^{(10)}_N}\int_{S^6}d^6 x\s{g}$. Consider
the following functional for any 2D surface $\gamma_A$ such that
$\de\gamma_A=A$ \be \f{1}{4G^{(10)}_N}\int_{\gamma_A\times
S^6}d^{8}x~ e^{-2\phi}\s{g}, \ee and try to minimize it. This
procedure singles out what should be called a minimal surface
$\gamma_A$. It is trivial to see that this procedure is reduced to
the original relation (\ref{arealaw}) when the dilaton is constant.

Then we find that $\gamma_A$ is defined by (the notation is the same
as in section \ref{EE for A_S}) \be
\f{dU}{dx}=\f{U^{5/2}}{g_{YM}\s{6\pi^2N}} \s{\f{U^7}{U^7_*}-1},\ee
where $U_*$ is the turning point of the surface and we assume
$g_{YM}^2N^{1/5}\ll U_*$. Following the same way of analysis in
section \ref{general results}, in the end we obtain the entanglement
entropy \ba S_{A_S}=\f{NLU_0^2}{5\pi g_{YM}^2} -c\cdot
\f{N^{5/3}L}{(g_{YM})^{2/3}l^{4/3}}, \label{resdt} \ea where $U_0$
is the UV cutoff (assuming $U_0\ll g_{YM}^2N$), and
$c=\f{1}{5}\left(\f{4\s{2}}{\s{3}}\right)^{4/3}\pi^{3/2}
\left(\f{\Gamma(5/7)}{\Gamma(3/14)}\right)^{7/3}$. The first term is
proportional to the length of $\gamma_A$ (i.e. $L$) in (\ref{resdt})
and is an analogue of the area law divergence term\footnote{It is
proportional to the square of the cut off energy and is different
from the area law relation (\ref{divarea}). However, this is not any
contradiction because we cannot set $U_0\to \infty$ due to the
constraint (\ref{constdt}). In such a high energy region, we cannot
neglect the stringy corrections and it is better to use the weakly
coupled Yang-Mills description.}. The second term is interesting
since it is finite and depends on $l$, non-trivially. Its $N$
dependence $\propto N^{5/3}$ is between the free field result $N^2$,
and the IR fixed point result $N^{3/2}$ (see (\ref{entompne})) of
the 3D ${\cal N}=8$ superconformal field theory, as expected. As in
the 4D case, we learn that the Yang-Mills interaction reduces the
degree of freedom.

\subsubsection{NS5-branes Case}

\hspace{5mm} The throat part of $N$ NS5-branes is described by the
following well-known metric \cite{CHS,IMSY}
 \ba ds^2&=&-dx^2_0+\sum_{i=1}^5 dx^2_i+N\al
\f{dU^2}{U^2}+N\al(d\Omega_3)^2,\no
e^{\phi}&=&\left(\f{(2\pi)^3N}{g_{YM}^2 U^{2}}\right)^{1/2}.
\label{sugrans} \ea We assume type IIB string theory to fix
notations. To take the decoupling limit, we keep the Yang-Mills
coupling $g_{YM}^2=(2\pi)^3\al$ finite and take the limit $g_s\to
0$. This leads to the little string theory (for a review see
\cite{aharony}). Notice that this theory is not a local field theory
and shows non-local behaviors such as the Hagedorn transition.

The calculation of the entanglement entropy can be done as before.
However, in this case\footnote{Via the S-duality the analysis of the
D5-branes leads to the same result.} of NS5-branes, we encounter the
following unusual feature. Consider the straight line case and try
to find solutions for the minimal surface equation \be
\f{dU}{dx}=\s{\f{(2\pi)^3}{Ng_{YM}^2}U^2\left(\f{U^4}{U_*^4}-1\right)}.
\ee Smooth solutions are allowed only for a fixed value of $l_*$ \be
l_*=\int^{\infty}_{U_*}
\left(\f{dx}{dU}\right)dU=\f{\s{Ng_{YM}^2}}{4\s{2\pi}}=\f{\pi}{2}\s{N\al}.
\label{smooth}\ee This suggests a phase transition at the energy
scale $(l_*)^{-1}$. Indeed, the value $l^{-1}\sim \f{1}{\s{N\al}}$
is the order of the Hagedorn temperature $T_H$ in the little string
theory\footnote{The holographic entanglement entropy in this case
takes the form of $S_{A_S}=c_1\cdot \f{N L^4}{g_{YM^2}}U_0^2-c_2L^4
N^2\f{U_*^2}{l^2}$, where $c_1$ and $c_2$ are a certain constant.}.
At least, we can claim from the computation in (\ref{smooth}) that
there is no solution when $l<l_*$. We can understand this because
the lack of locality means that we cannot define the entanglement
entropy when the size of $A$ becomes the same order of $T_H^{-1}$.

\section{Conclusions and Discussions}
\label{conclusion} \setcounter{equation}{0} \hspace{5mm} In this
paper we presented detailed discussions of the
holographic interpretation of the entanglement entropy proposed in
our earlier letter \cite{RuTa}. We gave a derivation of our proposal
(\ref{arealaw}) in the AdS$_3/$CFT$_2$ case by applying the basic
computation \cite{ADSGKP,ADSWitten} of correlation functions in
AdS/CFT correspondence. As for the higher dimensional case, we are
still lacking its complete derivation from standard AdS/CFT
correspondence, even though we offered an intuitive explanation and
several non-trivial evidences of our proposal (\ref{arealaw}). This
deserves further investigations. The proof of the strong
subadditivity (\ref{exttt}) will also be a non-trivial test for this
purpose, say.

The application of the proposal (\ref{arealaw}) to various quantum
field theories is also intriguing. Since we mainly analyzed the
conformal field theories, it would be useful to compute the
entanglement entropy in massive theories. In this paper, we did a
rough approximation by cutting off the IR region by hand and also
analyzed simple non-conformal backgrounds of D2-branes and
NS5-branes. The next step will be to compute the entanglement
entropy by considering supergravity backgrounds dual to more
realistic massive theories such as 4D confining theories. There we
expect that the entanglement entropy can be used as an alternative
of the Wilson loop to distinguish the confinement. Indeed, we
already noticed that the behavior of entanglement entropy is
drastically changed before and after the deconfinement phase
transition in ${\mathcal{N}}=4$ super Yang-Mills theory at finite
temperature from the AdS$_5$ side. Also, we obtained a singular
behavior of the entanglement entropy in the background with many
NS5-branes, which will probably be related to the non-locality or
the Hagedorn transition in the little string theory.

We also investigated the properties of the entanglement entropy from
the conformal field theory side. Especially we showed that important
parts of entanglement entropy are proportional to central charges in
any 4D CFTs from the analysis of Weyl anomaly. Even though we did
not find this property for the other parts of the entropy, which are
invariant under the Weyl scaling, the holographic analysis tells us
that the total entanglement entropy in strongly coupled 4D CFTs is
proportional to the central charge $a$.
These facts offer us an evidence that the central charge is
proportional to degrees of freedom in a given conformal field
theory. It would be an interesting future problem to study the
relation between possible $c$-theorems in more than two dimensions and
the property of entanglement entropy.

Several aspects of the entanglement entropy revealed in this paper
can have many implications on (strongly interacting) QFTs, some of
which might be realized in condensed matter physics, say. For
example, we derived the scaling of entanglement entropy
(\ref{areatwo}) for a compact submanifold $A$ based on AdS/CFT
correspondence where the coefficients $p_d$ and $q$ are universal
and conformal invariant. We expect that this is a generic feature
which might be applicable for systems that does not necessarily have
gravity (AdS) description. Thus, it is interesting to investigate
these quantities in several strongly interacting systems at
criticality. In a sense, these quantities are a generalization of
the central charges in CFTs in even spacetime dimensions, or the
quantum dimension in topological field theories. (Note also that
there is no counter part of the central charges in odd spacetime
dimensions.) For example, at least in principle, we can numerically
study these universal quantities in the entanglement entropy in
gapless spin liquids, and compare them with those computed from
several candidate effective field theories \cite{Wen89}. Also, even
though these effective field theories are suspected to be a gauge
theory, it might not be straightforward to identify the Wilson loop
operator in a generic microscopic spin model. In that situation, one
can instead look at the entanglement entropy since our analysis for
AdS black holes suggests it can be at least as useful as the Wilson
loop.

Finally, our computation of entanglement entropy may also be useful
to uncover holographical duals of string theory backgrounds which
are not well-understood, such as de-Sitter spaces and G\"{o}del
spaces\footnote{Recent discussions on these spaces from this
viewpoint can be found e.g. in \cite{desitter} and \cite{Godel}.}.
This is because the entanglement entropy captures the basic degrees
of freedom in the dual theory and because it can be easily estimated
classically in the gravity side.

\vskip6mm
\noindent
{\bf Acknowledgments}

\vskip2mm We are very grateful to M. Einhorn, D. Fursaev, A.
Kapustin, M. Levin, A. W. W. Ludwig, T. Okuda,  J. Preskill, and
especially to H. Ooguri and M. Shigemori for useful discussions and
to R. Emparan for helpful correspondence. The work was supported in
part by the National Science Foundation under Grant No.\
PHY99-07949.

\noindent


\newcommand{\J}[4]{{\sl #1} {\bf #2} (#3) #4}
\newcommand{\andJ}[3]{{\bf #1} (#2) #3}
\newcommand{\AP}{Ann.\ Phys.\ (N.Y.)}
\newcommand{\MPL}{Mod.\ Phys.\ Lett.}
\newcommand{\NP}{Nucl.\ Phys.}
\newcommand{\PL}{Phys.\ Lett.}
\newcommand{\PR}{ Phys.\ Rev.}
\newcommand{\PRL}{Phys.\ Rev.\ Lett.}
\newcommand{\PTP}{Prog.\ Theor.\ Phys.}
\newcommand{\hep}[1]{{\tt hep-th/{#1}}}

\end{document}